\documentclass{aa}

\usepackage{graphics}
\usepackage{epsfig}

\def\bib{\bibitem{}}

\newcommand{\xia}{\overline{\xi}}
\newcommand{\rhob}{\overline{\rho}}
\newcommand{\rhoa}{\overline{\rho}}
\newcommand{\gam}{\gamma}

\newcommand{\pl}{\partial}
\newcommand{\beq}{\begin{equation}}
\newcommand{\eeq}{\end{equation}}
\newcommand{\lag}{\langle}
\newcommand{\rag}{\rangle}
\newcommand{\Om}{\Omega_{\rm m}}
\newcommand{\Ol}{\Omega_{\Lambda}}
\newcommand{\Ob}{\Omega_{\rm b}}
\newcommand{\fEd}{f_{\rm Ed}}
\newcommand{\LEd}{L_{\rm Ed}}
\newcommand{\x}{u}
\begin{document}
%
%
%
%
\thesaurus{Sect.02 (11.03.1; 11.05.2; 11.12.2; 11.17.3; 12.12.1; 12.03.1)}
\title{Multiplicity Functions and X-ray Emission of Clusters and Groups versus  
Galaxies and Quasars}   
\author{P. Valageas \and R. Schaeffer}  
\institute{Service de Physique Th\'eorique, CEN Saclay, 91191 Gif-sur-Yvette, 
France} 
\date{Received 17 September 1999 / Accepted 26 April 2000}
\maketitle
\markboth{Valageas \& Schaeffer: Multiplicity Functions and X-ray Emission of 
Clusters and Groups versus Galaxies and Quasars}{Valageas \& Schaeffer: 
Multiplicity Functions and X-ray Emission of Clusters and Groups versus Galaxies 
and Quasars}

\begin{abstract}

We use a unified analytical formulation - developed in previous papers - for 
the multiplicity functions of clusters and galaxies. This method is free from the cloud-in-cloud problem encountered in earlier approaches and well adapted to the description of the non-linear clustering features. It is especially suited to {\it simultaneously} describe rich clusters, groups and galaxies, consistently with the hierarchical picture of gravitational clustering, as well as their evolution in time. We find a good agreement with observations for the cluster temperature function and we compare our method with the standard Press-Schechter prescription. We also obtain the main properties of the Sunyaev-Zel'dovich effect (mean and variance). Then, using a simple model for the cluster X-ray luminosity (taking into account entropy considerations), we obtain the X-ray luminosity distribution of groups and clusters. 

Then, using the same formalism we derive the galaxy and quasar multiplicity 
functions. In particular, we show that the use of the standard Press-Schechter 
prescription leads to erroneous conclusions at low redshifts while our 
approach provides a reasonable agreement with observations in a natural fashion 
because it is able to distinguish galactic halos from groups or clusters. 
Finally, we derive the contribution of quasars and galaxies to the X-ray counts. 
Thus, we obtain {\it a global and consistent picture of the X-ray emissions from 
all structures}. In particular, we show that future observations (e.g., from 
AXAF) could provide interesting information on galaxy evolution. Indeed, they 
will constrain the importance of a possible hot diffuse gaseous phase in 
galactic halos and they could reveal massive galaxies which are just being 
formed, through the X-ray emission of their cooling gas.

\end{abstract}

\keywords{galaxies: clusters - galaxies: evolution - galaxies: mass function - 
quasars: general - cosmology: large-scale structure of Universe - cosmic 
microwave background}

\section{Introduction}

Clusters of galaxies are the largest virialized objects, characterized by scales 
which have just entered the non-linear regime. Thus, they are very rare and 
their number density is very sensitive to the cosmological parameters and to the 
amplitude of the initial density fluctuations (within the framework of the usual 
hierarchical scenarios). As a consequence, as proposed by Oukbir \& Blanchard 
(1992) many authors have compared observations with predictions from numerical 
simulations (e.g., Eke et al. 1996) or analytic approaches (e.g., Oukbir \& 
Blanchard 1997) in order to obtain constraints on $(\Om,\Ol)$ and on the initial 
power-spectrum $P(k)$. However, the cluster temperature - X-ray luminosity 
relation involves non-gravitational effects since simple scaling laws (Kaiser 
1986) recovered by numerical simulations (Eke et al. 1998) that  neglect 
radiative cooling and supernova or quasar feedback disagree with observations. 
This could be due to a preheating of the IGM by QSOs or supernovae (Valageas \& 
Silk 1999b; Cavaliere et al. 1997; Ponman et al. 1999) which has not been considered yet 
by numerical simulations. Thus, analytic approaches are still needed in order to 
describe clusters. Moreover, they explicitly show the connection of cluster 
characteristics (e.g., their mass function) with other features of the universe 
(e.g., various properties of the underlying density field, of galaxies or 
quasars). 

The standard way to estimate the mass multiplicity function 
of collapsed objects by analytical means, is the well-known Press-Schechter (1974) approximation 
- hereafter PS - that directly recognizes 
 in the initial, linear, fluctuations of the density field
the overdensities that will eventually 
collapse. This 
approximation, however, suffers from many drawbacks. One is that it is customary 
-and necessary in order to bring the analytical form in agreement with numerical 
simulations- to multiply the result by an ad-hoc factor of two. This cannot be 
justified by the standard excursion sets argument for realistic filters like the 
top-hat in real-space as shown in Valageas \& Schaeffer (1997) - hereafter VS - 
(see also Peacock \& Heavens 1990). Indeed, in such a case the excursion sets 
imply a renormalization factor which goes to unity at large masses. 

Another fundamental problem, inherently related to hierarchical clustering, is 
to describe objects embedded within other, less dense but nevertheless 
virialized, objects, that is to describe {\it subclustering}. This has long 
been known (Bardeen et al. 1986) not to be reliable within approaches directly 
based on recognition of the linear overdensities that will form objects, and 
is called the {\it cloud-in-cloud problem}. Indeed, an approach based on 
counting overdensities in the linear regime leads to severe overcounting, so 
severe that the same objects are erroneously counted an infinite number of times (VS). 
Thus, to describe a universe made of dense galaxies ($10^4$ times the mean 
density), as well as of galaxy clusters ($10^2$ times the mean density), is out of 
reach of such approaches, while the description of intermediate objects such as 
Ly$\alpha $ clouds, that may even have densities below the mean, is unthinkable 
within this framework. Indeed, note that the PS approximation can only deal with 
just-virialized halos.

Recent progress (VS) in the description of the non-linear density field, through 
a {\it non-linear scaling model} based on earlier approaches (Schaeffer 1984,
1985; 
Balian \& Schaeffer 1989; Bernardeau \& Schaeffer 1991), with an understanding 
of its relation to the initial spectrum of fluctuations, leads to analytical 
expressions that enable one to  {\it directly count the overdensities in the 
actual non-linear density field}. The result exhibits many analogies with the PS 
approximation, but it gives a correct normalization and it also solves the 
cloud-in-cloud problem (there are no divergences). It also provides the 
correlations of these objects and the associated bias as compared to the matter 
distribution (Valageas et al. 2000b; Bernardeau \& Schaeffer 1992,1999).

For the purposes of the present paper, the scaling approach gives a clear answer 
to the following problem. Choose {\it any} density contrast $\Delta = 
(\rho-\rhob) / \rhob$ and define objects as having an inner density contrast 
larger than this threshold. Then {\it all} the mass in the universe (except a 
nonvanishing but negligible fraction in the very underdense regions) lies within 
such objects, that have a distribution of mass or size given by the theory. The 
choice of a larger contrast results in a different partition of the same mass 
into smaller objects, the procedure being valid for arbitrarily large contrasts 
(with obvious limits at the kpc scale where pressure and angular momentum 
come into play)
 as long as one remains in the non-linear regime. This is the way {\it 
subclustering} can be properly accounted for. Then, the same theory can 
simultaneously describe virialized clusters, which have a contrast of $\simeq 
200$, as well as galaxies (Valageas \& Schaeffer 1999), whose contrast is 
$\simeq 5000$, embedded or not in the latter objects. The procedure also holds for small density contrasts, even negative, provided one is at scales (in the 
present universe, for instance, not much above 1 Mpc) where the correlation 
function is large enough so as to insure one is in the fully non-linear regime. 
This corresponds to {\it underdense non-linear objects} embedded in much  less 
dense, nearly void regions. The new feature introduced by the scaling model is 
that, because one directly works in the non-linear regime, this separation can 
be done properly. The same procedure applied to future non-linear objects in the 
linear regime as is done in the PS approach would lead to the above mentioned 
divergences (VS), which is another indication that defining objects directly in 
the linear regime, as is done in the latter approach, is unsecure.

The analytical predictions of the scaling model have been checked against 
numerical simulations (Valageas, Lacey \& Schaeffer 2000a) under the most 
extreme conditions, searching for non-linear objects with density contrasts 
ranging from values as large as $\Delta = 5000$ down to negative values $\Delta 
= -0.5$, spanning four orders of magnitude and limited only by the accuracy of 
the simulations. Thus, this approach provides reasonable 
results, at least in the above range, allowing us to describe non linear objects 
and their evolution. These theoretical predictions are also in the line of the 
findings of Moore et al. (1999a) that resolve some of the structure of dark matter 
halos in their simulation. As a consequence, the mass functions provided by this 
approach give, with the same parameters, {\it a unified description of very 
different objects}, such as galaxies and quasars (Valageas \& Schaeffer 1999), 
Ly$\alpha$ clouds (Valageas, Schaeffer \& Silk 1999), the ionization flux 
emitted by QSOs (Valageas \& Silk 1999a), allowing for a consistent picture of 
the reheating and the reionization history of the universe (Valageas \& Silk 
1999a,b). Although this had been checked from the beginning of this series of 
articles, it remains to be explicitly shown that the same approach, without new 
parameters, provides the correct cluster multiplicity and its observed evolution 
with redshift.

Thus, the main goals of this article are to:

- check the predictions for the properties of clusters of an analytic method 
developed in previous studies which can describe in a consistent way the mass 
functions of various objects (clusters, galaxies, Ly$\alpha$ clouds, etc.) while 
making the connection with other properties of the density field (correlation 
functions, counts-in-cells, etc.).

- evaluate by means of the same method the distortion of the CMB anisotropies 
induced by these X-ray clusters through the Sunyaev-Zel'dovich effect.

- introduce a simple model which can reproduce the observed temperature - X-ray 
luminosity relation of groups and clusters.

- use the global scope of our description to compare the X-ray emission provided 
by clusters, groups, galaxies and quasars. In particular, we show that this 
requires us to go beyond the standard PS prescription in order to deal with various 
classes of objects which may not be defined by the usual constant density 
threshold $\Delta_c(z) \sim 177$.

In Sect.\ref{Multiplicity functions}, we present the expressions of the 
multiplicity functions and we compare the PS approach with the scaling model, 
focussing on the cluster mass function. The temperature function and its 
evolution with redshift are discussed in Sect.\ref{Cluster temperature 
functions} while Sect.\ref{Sunyaev-Zel'dovich effect} is devoted to the 
implications of our model for the characteristics of the Sunyaev-Zel'dovich 
effect. Next, in Sect.\ref{Evolution of the X-ray luminosity function} we 
present a model for the temperature - X-ray luminosity relation and we derive 
the cluster X-ray luminosity function. Finally, in Sect.\ref{Galaxies and 
quasars versus groups and clusters} we describe the galaxy and quasar luminosity 
functions predicted by our model, which allows us to draw a complete picture of 
the X-ray emission from all structures.

\section{Multiplicity functions}
\label{Multiplicity functions}

\subsection{Formalism}
\label{Formalism}

We define clusters as halos with a mean density contrast equal to $\Delta_c$, 
where $\Delta_c$ is the density contrast at the time of virialization, at the 
redshift $z$ we consider, given by the spherical collapse model. This means that 
clusters just virialize at the redshift at which we see them. For a critical 
density universe this density threshold is a constant: $\Delta_c \simeq 177$. 
Numerical simulations show that this value of the density contrast separates 
reasonably well the virialized halos from the surrounding material still falling 
onto the overdensity (Cole \& Lacey 1996). This justifies this traditional 
definition of clusters.

Following the method outlined in VS we shall use two prescriptions to get the 
comoving mass function of these halos. Firstly, we recall the {\it scaling 
model} developed in previous papers, which we apply here to clusters of 
galaxies. Secondly, we also consider the usual PS approximation for the sake of 
comparison. Undoubtedly, we have in mind that the former supercedes the latter 
in the sense that it takes full benefit of the hierarchical clustering picture. 
The main advantages of this scaling approach are i) to make the link between the 
mass functions and the counts-in-cells statistics and ii) to provide a very 
powerful tool which can describe many different mass functions (i.e. defined by 
various density thresholds) as well as other properties of the non-linear 
density field.

The scaling model assumes that the many-body correlation functions follow 
specific scaling laws obtained from the stable-clustering ansatz (see VS for 
details). Then, we attach to each object a parameter $x$ defined by:
\beq
x(M,z) =  \frac{1+\Delta_c}{\; \xia[R(M,z),z] \;} ,
\label{xnl}
\eeq
where 
\[
\xia(R) =   \int_V \frac{d^3r_1 \; d^3r_2}{V^2} \; \xi_2 ({\bf r}_1,{\bf r}_2) 
 \;\;\;\;\; \mbox{with} \;\;\;\;\; V= \frac{4}{3} \pi R^3
\]
is the average of the two-body correlation function $\xi_2 ({\bf r}_1,{\bf 
r}_2)$ over a spherical cell of radius $R$ and provides the measure of the 
typical density fluctuations in such a cell. Thus large $x$ correspond to 
deep, and small $x$ to shallow potential wells. Then, we write the 
multiplicity function of these objects,  defined by the density 
threshold $\Delta_c$, as (see VS):
\beq
\eta(M) \frac{dM}{M}  = \frac{\rhoa_0}{M} \; x^2 H(x) \; \frac{dx}{x} ,
\label{etah}
\eeq
where $\rhoa_0$ is the mean density of the universe at $z=0$. The scaling 
function $H(x)$ only depends on the initial spectrum of the density fluctuations 
and must be obtained from numerical simulations. However, from theoretical 
arguments (see VS, Bernardeau \& Schaeffer 1992 and Balian \& Schaeffer 1989) it 
is expected to follow the asymptotic behaviour:
\[
x \ll 1 \; : \; H(x) \propto x^{\omega-2} \hspace{0.3cm} , \hspace{0.3cm} 
x \gg 1 \; : \; H(x) \propto x^{\omega_s-1} \; e^{-x/x_*}
\]
with $\omega \simeq 0.5$, $\omega_s \sim -3/2$, $x_* \sim 10$ to 20 and by 
definition it must satisfy:
\beq
\int_0^{\infty}  x \; H(x) \; dx   =  1 .
\eeq
This formulation is directly linked to the statistics of the counts-in-cells 
which involve a scaling function $h(x)$, related to the probability 
$P(\Delta,R)$ to have a density contrast $\Delta$ in a spherical cell of fixed 
radius $R$, that scales as:
\beq
(1+\Delta) \;  P(\Delta  ,R) \;  d \Delta  = x^2\;h(x) \frac{dx}{x}   \;\;\; , 
\;\;\; 
x(\Delta)  =  \frac{1+\Delta}{ \xia[R,z] } .
\eeq
This function $h(x)$ is very close to $H(x)$, see VS and Valageas et al. (2000a) 
for details. The relevance of these scaling functions
 has been checked using  numerical simulations for CDM initial 
conditions by Bouchet et al. (1991) and, more systematically using power-law 
initial spectra in Colombi et al. (1997), Munshi et al. (1999) and Valageas et 
al. (2000a). Here we note that in principle this prescription 
only applies to the highly non-linear 
regime ($\xia \ga  100$) while clusters correspond to mildly non-linear scales 
($\xia \sim 10 - 100$). Hence the cluster mass functions we obtain may show an 
accuracy of $~10\%$. Note however that the scaling function we shall use, taken 
from numerical simulations by Bouchet et al. (1991), was measured in this range 
of $\xia$, and was indeed found to describe clustering at the above accuracy.

On the other hand, the usual PS approximation gives
\beq
\eta(M) \frac{dM}{M} = \sqrt{\frac{2}{\pi}} \frac{\rhoa_0}{M} 
\frac{\delta_c}{\sigma} \left| \frac{\mbox{dln}\sigma}{\mbox{dln}M} \right| \exp 
\left[ - \frac{\delta_c^2}{2\sigma^2} \right] \frac{dM}{M}  , 
\label{etaPS}
\eeq
where $\delta_c(z)$ is the present linear density contrast for halos which 
collapsed at redshift $z$ according to the spherical collapse model. As usual, 
$\sigma(M)$ is the rms density fluctuation extrapolated by linear theory at 
$z=0$ at scale $M$. Here we multiplied the mass function (\ref{etaPS}) by the 
usual empirical factor of 2 so that all the mass is contained in such 
overdensities.

We can see that {\it the scaling mass function} (\ref{etah}) {\it predicts more 
numerous very massive halos but fewer small objects than the PS mass function} 
(\ref{etaPS}). This difference can be directly seen through the scaling 
function $H(x)$. Indeed, as shown in VS the PS formulation can be translated 
into the scaling approach in the highly non-linear regime where it leads to a 
specific scaling function:
\beq
h_{PS}(x) = \sqrt{\frac{2}{\pi}} \frac{5+n}{6\alpha} x^{\frac{5+n}{6} -2} \exp 
\left[ - x^{(5+n)/3} /(2\alpha^2) \right] 
\label{hPS}
\eeq 
where $n$ is the slope of the power-spectrum and $\alpha \simeq 1$. A comparison 
with the function $h(x)$ directly measured in numerical simulations from 
counts-in-cells (see VS and Valageas et al. 2000a) shows that the large-mass 
cutoff of the PS mass function is too sharp while its peak (at masses $~M_*$ 
where $\sigma(M_*)=1$) is too high. This latter region corresponds to low mass 
clusters in Fig.\ref{figM} and to low temperature objects in Fig.\ref{figTO1} 
and Fig.\ref{figTO03} below. Note however that at very low masses, corresponding 
to very small galaxies, the PS mass functions give fewer objects than the 
scaling prescription as seen in Valageas \& Schaeffer (1999).

Since in this article we wish to obtain in a consistent fashion the X-ray 
emission from all discrete sources: clusters, quasars and galaxies, we also need 
the multiplicity function of galaxies. As described in Valageas \& Schaeffer 
(1999) we define galaxies by {\it two} constraints
to be satisfied simultaneously: the usual density threshold 
$\Delta_c(z)$ (as for clusters) {\it and } a cooling condition which ensures 
that the gas is able to dissipate its energy and form stars. This implies that 
galactic halos are defined by a density contrast $\Delta_{gal}(x,z)$ which 
depends on the mass of the object. Then, we can still use the expressions 
(\ref{xnl}) and (\ref{etah}) to obtain the galaxy mass function, where the 
density contrast is now set to $\Delta_{gal}(x,z)$ and depends on $x$ (see VS 
and Valageas et al. 2000a). Moreover, from the identification (\ref{hPS}) we can 
also consider an ``extended PS'' prescription to count these halos. We shall 
use this below in Sect.\ref{Galaxies and quasars versus groups and clusters} 
when we deal with galaxies and quasars.

For the numerical calculations we shall consider two cosmologies. First, we 
study a critical density universe (SCDM) with $\Ob=0.04$, $H_0=60$ km/s, 
$\sigma_8=0.5$ and a CDM power-spectrum (Davis et al. 1985). Next, we consider an 
open CDM universe (OCDM) with $\Om=0.3$, $\Ol=0$, $\Ob=0.03$, $H_0=60$ km/s and 
$\sigma_8=0.77$. These values are those we used in previous articles where we 
considered the luminosity functions of galaxies (Valageas \& Schaeffer 1999), 
Lyman-$\alpha$ absorbers (Valageas et al. 1999) and reionization by stars and 
quasars (Valageas \& Silk 1999a). Thus, this present study of clusters and 
groups of galaxies completes our description of structure formation in the 
universe, so that we obtain a unified model which can describe in a consistent 
fashion all these objects, from small low-density Lyman-$\alpha$ absorbers up to 
massive clusters.
 
Recent CMB observations (e.g., de Bernardis et al. 2000) suggest that the universe is flat. However, for our purposes, a $\Lambda$CDM universe with $\Om=0.3$ and $\Ol=0.7$ should be very close to our OCDM scenario, with a slight modification of the parameters of our astrophysical model for galaxies and clusters. In particular, as seen in Peacock \& Dodds (1996) the functional relation between the linear and non-linear power-spectra is the same for the OCDM and $\Lambda$CDM (i.e. one simply needs to take into account the variation of the linear growth factor). Moreover, the scaling model which leads to the multiplicity function (\ref{etah}) has been seen to agree with the statistics of the projected density $\kappa$ along the line of sight obtained from numerical simulations for all three SCDM, OCDM and $\Lambda$CDM cosmologies, as shown for instance in Valageas (2000). These results strongly suggest that our approach could also be used for the $\Lambda$CDM case. (In a similar fashion, the accuracy of the PS prescription is similar for all three cosmologies).

\subsection{Evolution with redshift of the cluster mass function}
\label{Evolution of the mass function}

\begin{figure}[htb]

\centerline{\epsfxsize=8 cm \epsfysize=5.5 cm \epsfbox{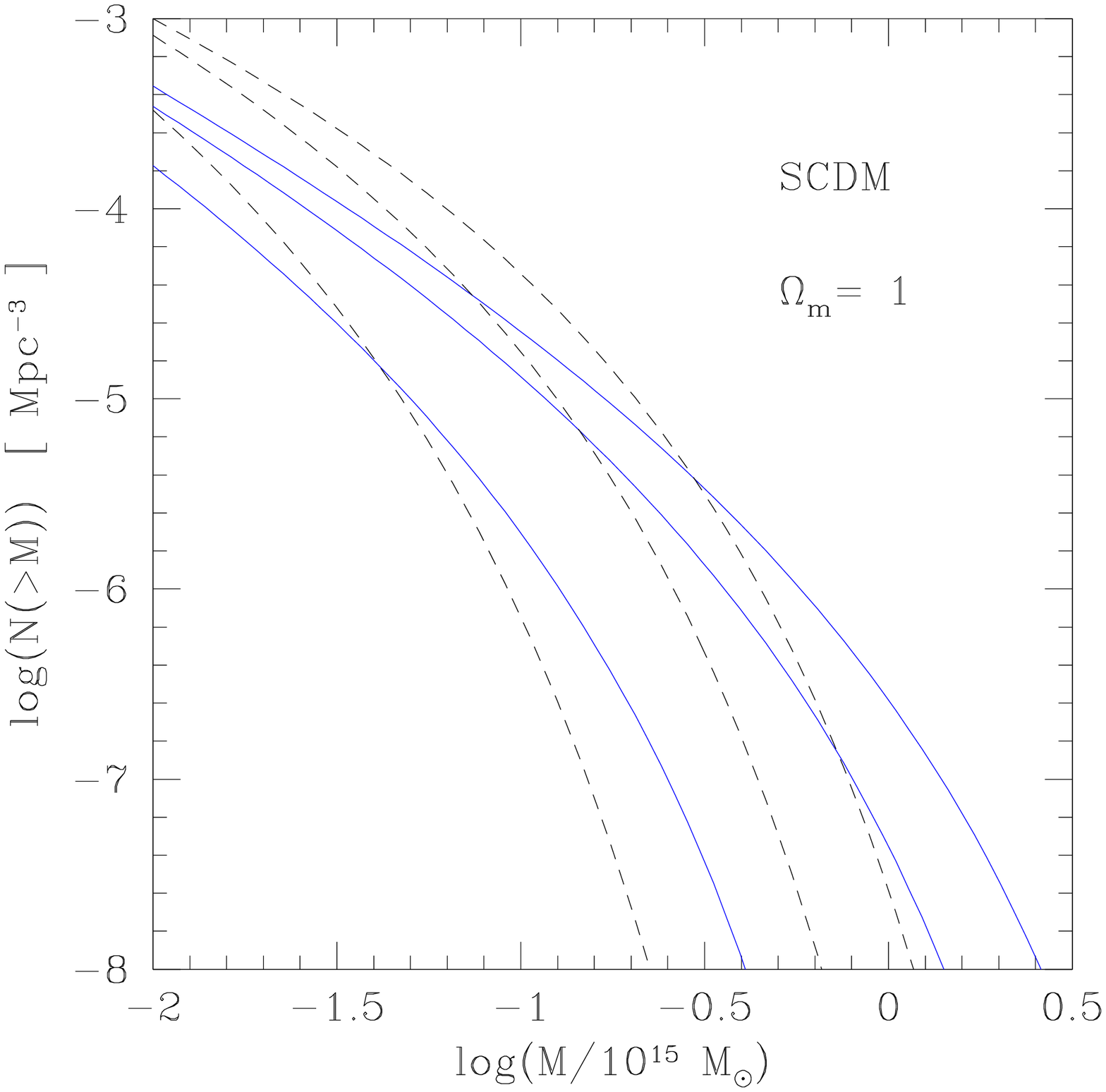}}
\centerline{\epsfxsize=8 cm \epsfysize=5.5 cm \epsfbox{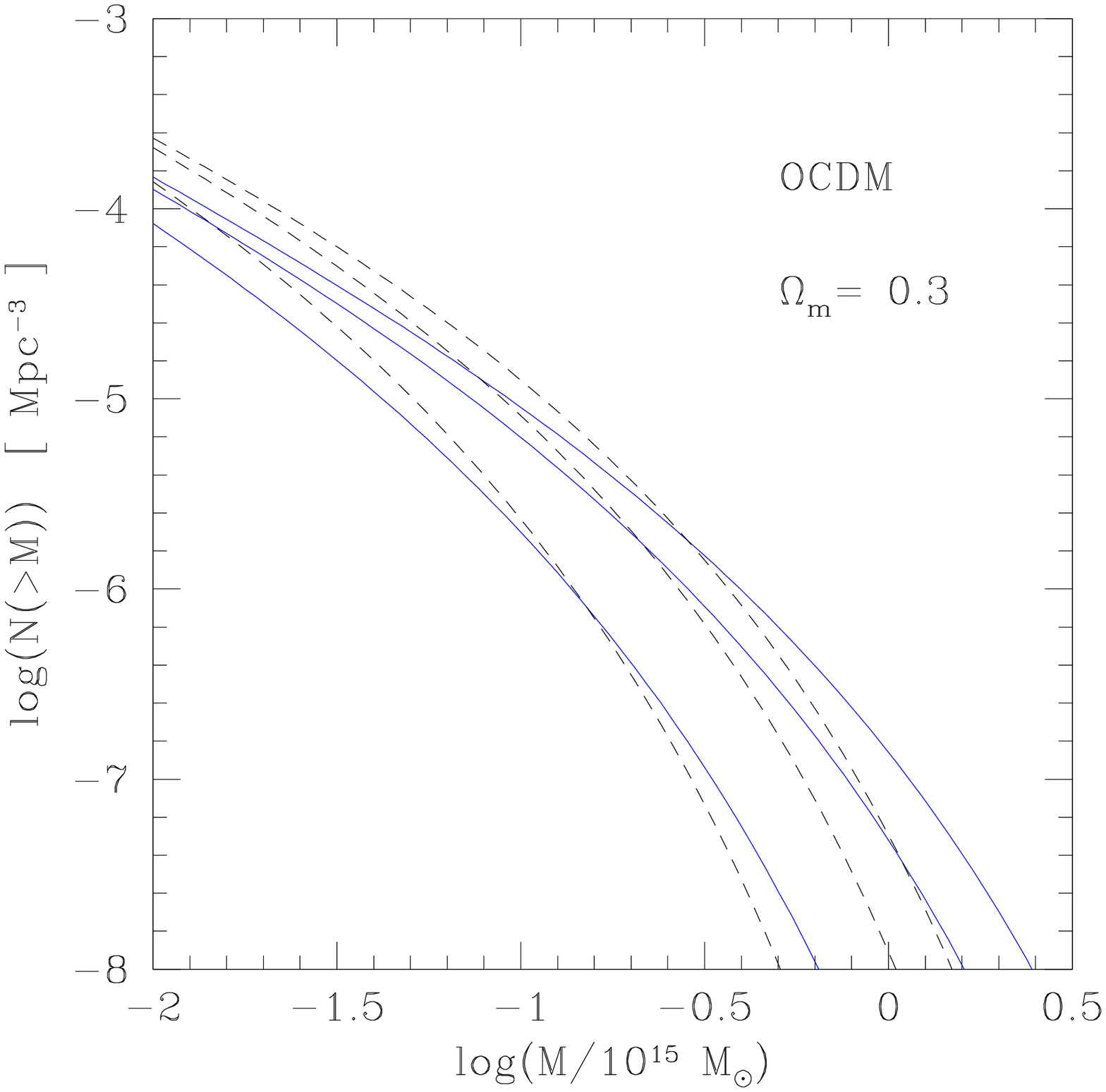}}

\caption{{\it Upper panel}: the cluster comoving cumulative mass function at 
redshifts $z=0.05, \; z=0.33$ and $z=1$, for a critical density universe with a 
CDM power-spectrum. Higher redshifts lead to fewer massive clusters. The solid 
lines correspond to the scaling prescription while the dashed lines represent 
the PS formulation. {\it Lower panel}: same curves for an open universe with 
$\Om=0.3$.}
\label{figM}

\end{figure}

We display in Fig.\ref{figM} the cluster comoving cumulative mass function we 
obtain for the two cosmological scenarios we consider in this article, at 
redshifts $z=0.05$, $z=0.33$ and $z=1$. First, we can check that we recover the 
trend described in the previous section: the scaling approach predicts fewer low 
mass halos and more very massive objects than the PS prescription. In a similar 
fashion, Governato et al. (1999) find that the PS mass function overestimates the 
number of small clusters ($M < 2 \; 10^{14} M_{\odot}$, for $\sigma_8 = 0.7$) 
and underestimates the number of massive halos (see also Gross et al. 1998 for a 
similar trend). However, the discrepancy they measure is smaller than ours. This 
may be due to the fact that the power-spectrum we use (from Davis et al. 1985) is 
slightly different from theirs and to the use of our scaling model in a range of 
$\xia$ slightly beyond its range of validity: indeed very high masses correspond 
to large scales which are getting close to the linear regime where the scaling 
model is not valid. Nevertheless, the agreement of our predictions (already 
described in a more general context in VS), with the behaviour observed in the 
numerical simulations, is quite encouraging.

We can check that the redshift evolution of the mass function is very sensitive 
to the cosmological parameter $\Om$ (e.g., Oukbir \& Blanchard 1992; Eke et 
al. 1996). Indeed, the number of clusters declines faster with $z$ for the 
critical density universe than for the open model. This is simply due to the 
fact that in the latter case structures have nearly stopped growing since the 
redshift where $\Om$ became appreciably smaller than unity, whence a very weak 
evolution with redshift, while in the former case structures keep building under 
the action of gravity at all epochs.

\begin{figure}[htb]

\centerline{\epsfxsize=8 cm \epsfysize=5.5 cm \epsfbox{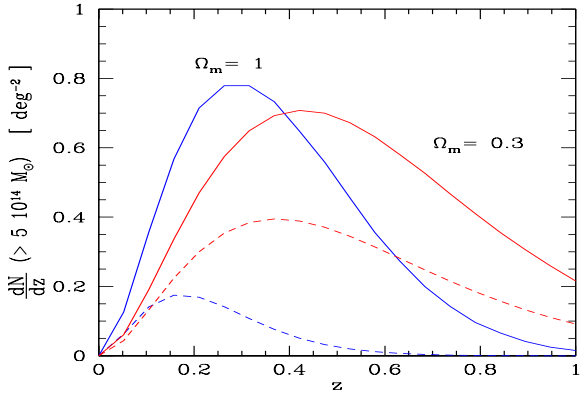}}

\caption{The redshift distribution per square degree of clusters more massive 
than $5 \times 10^{14}\;M_{\odot}$, for $\Om=1$ and $\Om=0.3$. The graph shows 
the number of clusters $d{\cal N}/dz(>5\;10^{14}\;M_{\odot})$ per unit redshift 
interval and square degree. The solid lines correspond to the non-linear 
prescription, while the dashed lines represent the PS formulation. The open 
universe corresponds to the slowest redshift evolution.}
\label{figNsupMz}

\end{figure}

The dependence on $\Om$ of the redshift evolution of the cluster mass function 
is even more apparent in Fig.\ref{figNsupMz} which shows the redshift 
distribution of clusters more massive than $5 \times 10^{14}\;M_{\odot}$, per 
square degree, for both SCDM and OCDM scenarios:
\beq
\!\!\!\! \frac{d{\cal N}}{dz} (>5 \times 10^{14} M_{\odot}) = \left( 
\frac{\pi}{180} \right)^2 \frac{dV}{d\Omega dz} \int_{5 \times 10^{14} 
M_{\odot}}^{\infty} \eta(M) \frac{dM}{M} 
\eeq
where $dV/d\Omega dz$ is the comoving volume element per unit steradian and unit 
redshift. The normalization of the PS prediction is lower than for the scaling 
approach as explained above (we count halos which are in the tail of the mass 
function). The main result of this figure is to emphasize the difference between the two  cosmologies  of the redshift evolution. Thus, for $\Om=1$ the number of such clusters reaches a maximum at $z \sim 0.3$, while for $\Om=0.3$ the peak 
corresponds to $z \sim 0.45$ and the evolution is slower.

\subsection{Counts along the line of sight}
\label{Line-of-sight}

\begin{figure}[htb]

\centerline{\epsfxsize=8 cm \epsfysize=5.5 cm \epsfbox{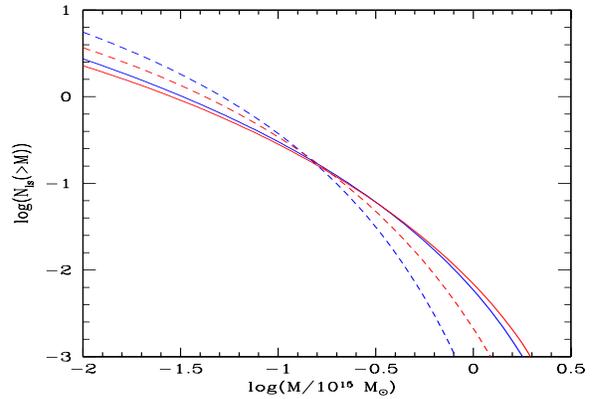}}

\caption{The mean number of clusters ${\cal N}_{ls}(>M)$ of mass larger than $M$ 
which intersect a line of sight between $z=0$ and $z=2$ (the contribution of 
higher redshifts is negligible). The solid lines correspond to the scaling 
prescription and the dashed lines to the PS approach, for both SCDM and OCDM 
scenarios. In both cases the number of very massive clusters is larger for the 
open universe.}
\label{figNls}

\end{figure}

From the cluster multiplicity function we can derive the mean number of clusters 
${\cal N}_{ls}(>M,<z)$ of mass larger than $M$ which intersect a line of sight 
between $z=0$ and a given redshift $z$:
\[
{\cal N}_{ls}(>M,<z) = \int_0^z c \frac{dt}{dz} (1+z)^3 dz \int_M^{\infty} 
\frac{dM}{M} \eta(M) \pi R^2 .
\]
The result is displayed in Fig.\ref{figNls}. Of course, we recover the difference 
between the scaling model and the PS approach we described above for the mass 
functions. The open universe gives higher counts for very massive halos because 
of the slower decline with redshift of the mass function. Note however that for 
the scaling prescription the difference between the two  cosmologies is quite small. 
Of course this would change with another choice for $\sigma_8$. We see that the 
mean number of clusters on a line-of-sight is quite small since we typically 
have ${\cal N}_{ls} < 1$.

\section{Cluster temperature function}
\label{Cluster temperature functions}

Although the study of cluster mass functions is convenient from a theoretical 
point of view, for observational purposes it is more interesting to consider 
temperature functions. For this we need the temperature which is
associated with halos of a given mass.

\subsection{Characteristic temperature of the halos}
\label{Characteristics of the halos}

The Jeans equation for the velocity dispersion $\sigma_v$ of the dark matter yields: 
\beq
\frac{d}{dr} (\rho \sigma_v^2) = - \; \rho \; \frac{{\cal G} M(<r)}{r^2} .
\label{Jeans}
\eeq
In the case of an isothermal density profile $\rho(r) \propto r^{-2}$ this leads to:
\beq
\sigma_v^2(R) = \frac{{\cal G} M}{2 R} \hspace{0.3cm} \mbox{and} \hspace{0.3cm} 
k T = \frac{{\cal G} \mu m_p M}{2 R} ,
\label{Tvir}
\eeq
where we defined the virial temperature $T$ of the halo by $k T = \mu m_p 
\sigma_v^2$ and $R$ is the virial radius of the cluster. Here $\mu m_p$ is the 
mean molecular weight of the gas and $m_p$ is the proton mass. For a different 
density profile we would still obtain (\ref{Tvir}) for the mean temperature, 
with a multiplicative factor of order unity. We assume that the mass of baryons 
$M_b$ is proportional to the mass of the dark matter halo $M$:
\beq
M_b = \frac{\Ob}{\Om} \; M ,
\eeq
where $\Ob$ is the present ratio of the baryon density to the critical 
density. With these parameters, using the fact that halos are defined by the 
density contrast $\Delta_c(z)$, we have:
\beq
M \propto \Om (1+\Delta_c) (1+z)^3 R^3
\label{MR}
\eeq
where $R$ is the virial radius of the cluster, and eq.(\ref{Tvir}) writes:
\beq
T = T_0 \; M_{15}^{2/3} \; \Delta_c(z)^{1/3} \; (1+z)  
\label{MT}
\eeq
with
\[
M_{15} = \left( \frac{M}{10^{15}\;M_{\odot}} \right) \;\;\; \mbox{and} \;\;\;   
T_0 = 1.2 \; \Om^{1/3} \; h^{2/3} \; \mbox{keV} .
\]
This is consistent with numerical simulations which recover this scaling law, 
with a similar normalization. In these units, Navarro et al. (1995) find $T_0 = 
1.4 \; \Om^{1/3} \; h^{2/3}$ keV while Evrard et al. (1996) get $T_0 = 1.2 
\; \Om^{1/3} \; h^{2/3}$ keV.

\subsection{Evolution with redshift of the temperature function}
\label{Evolution of the temperature function}

\begin{figure}[htb]

\centerline{\epsfxsize=8 cm \epsfysize=5.5 cm \epsfbox{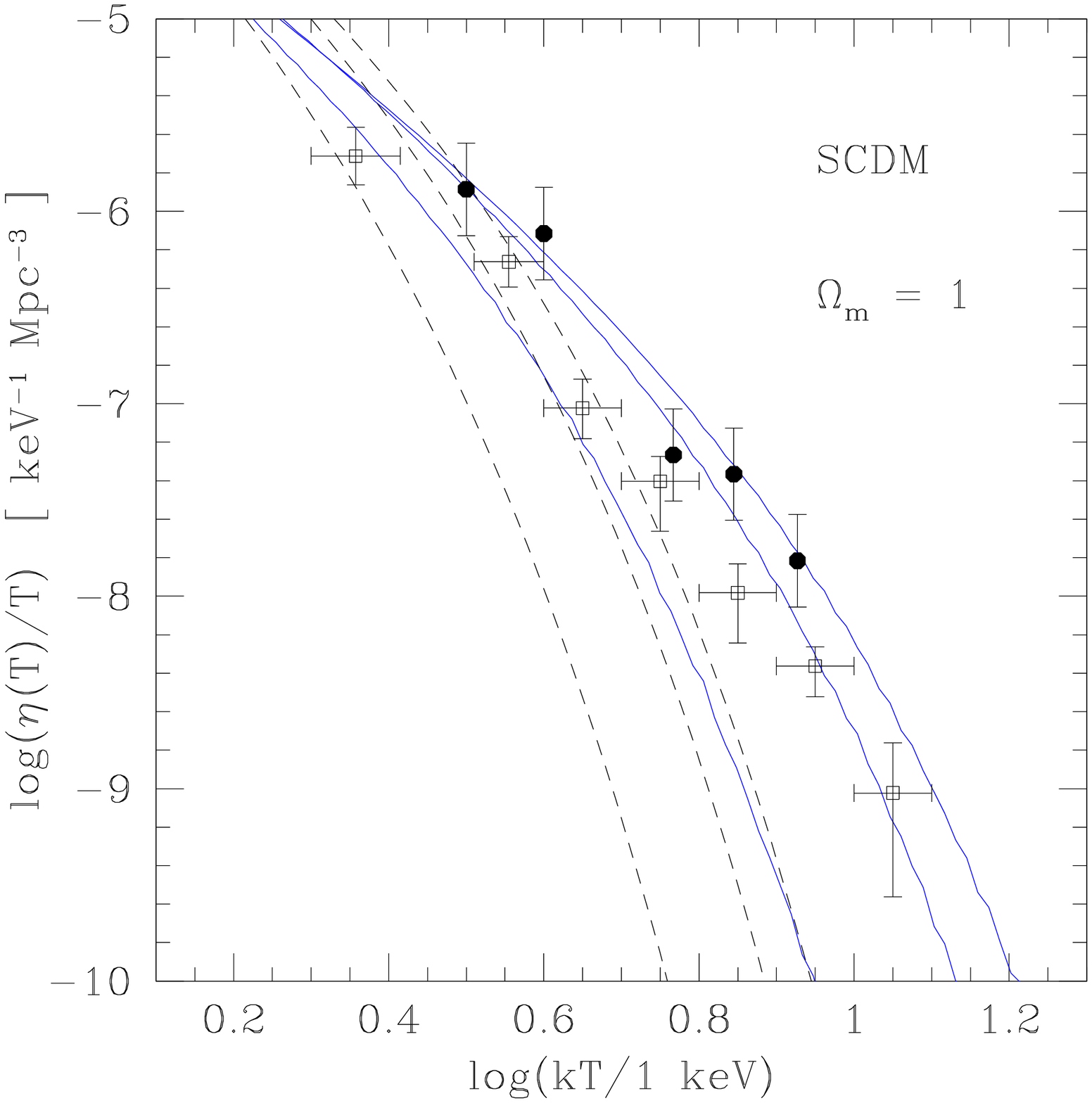}}
\centerline{\epsfxsize=8 cm \epsfysize=5.5 cm \epsfbox{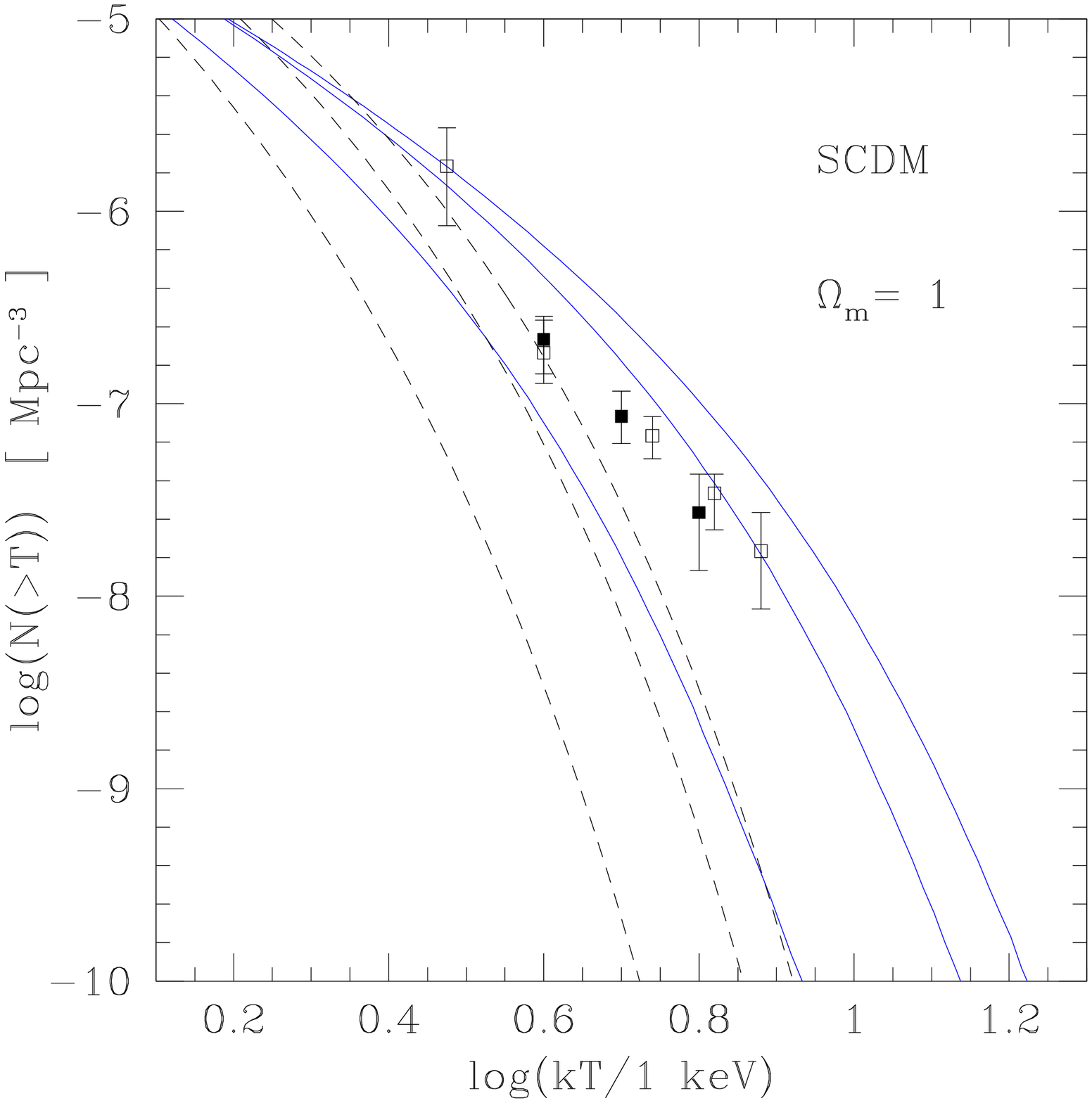}}

\caption{{\it Upper panel}: the comoving cluster X-ray temperature function 
$\eta(T) \; dT/T$ for the SCDM scenario. The solid lines correspond to the 
scaling prescription while the dashed lines represent the PS formulation. We 
display the redshifts $z=0.05$, $z=0.33$ and $z=1$ (a larger $z$ corresponds to 
fewer bright clusters). The data points are observational results at $z=0$ from 
Henry \& Arnaud (1991) (disks) and Edge et al. (1990) (squares). {\it Lower 
panel}: the cumulative cluster temperature function at the same redshifts. The 
open and filled squares are observations from Henry (1997) at $z=0.05$ and 
$z=0.33$ respectively.}
\label{figTO1}

\end{figure}

\begin{figure}[htb]

\centerline{\epsfxsize=8 cm \epsfysize=5.5 cm \epsfbox{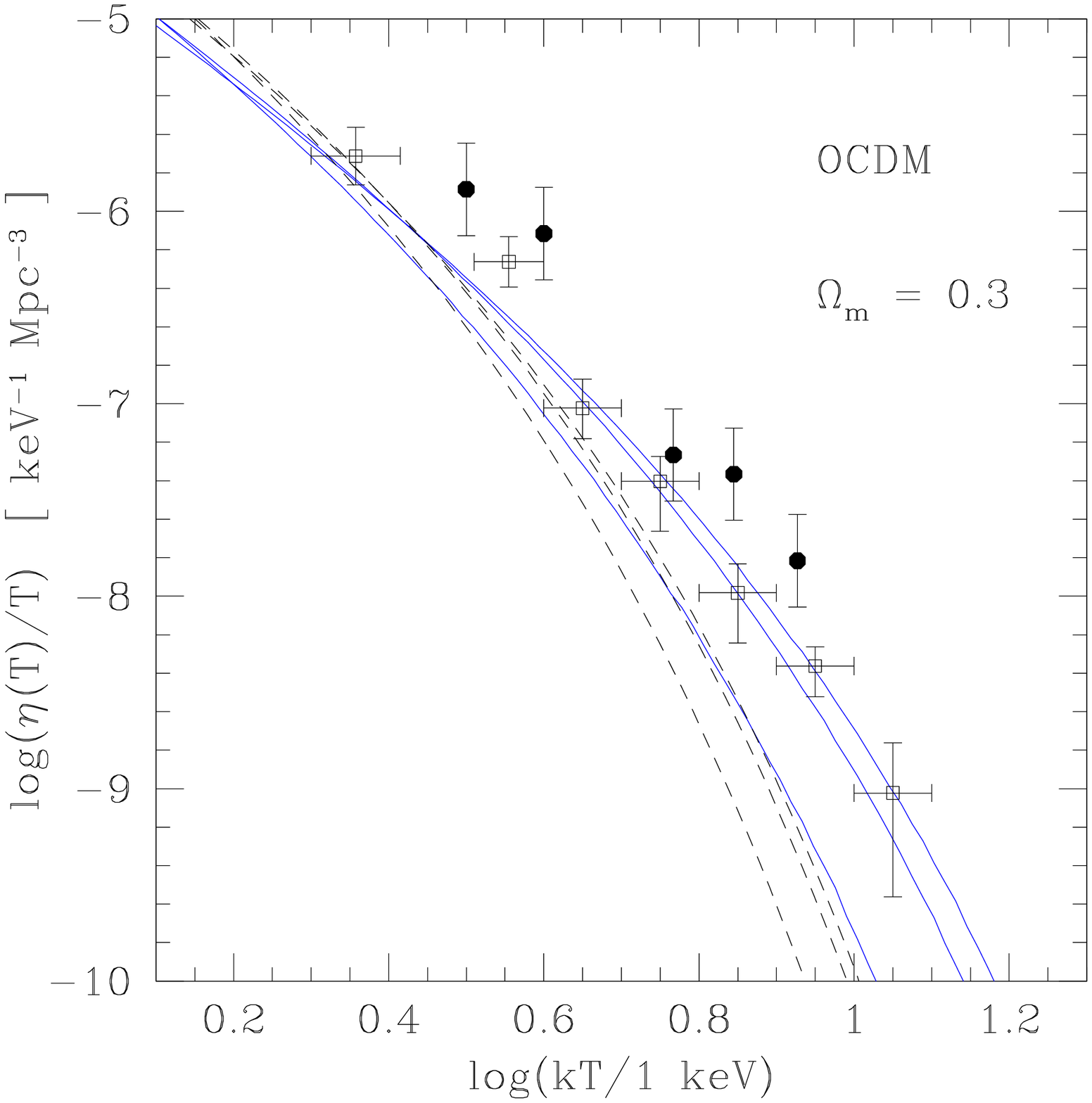}}
\centerline{\epsfxsize=8 cm \epsfysize=5.5 cm \epsfbox{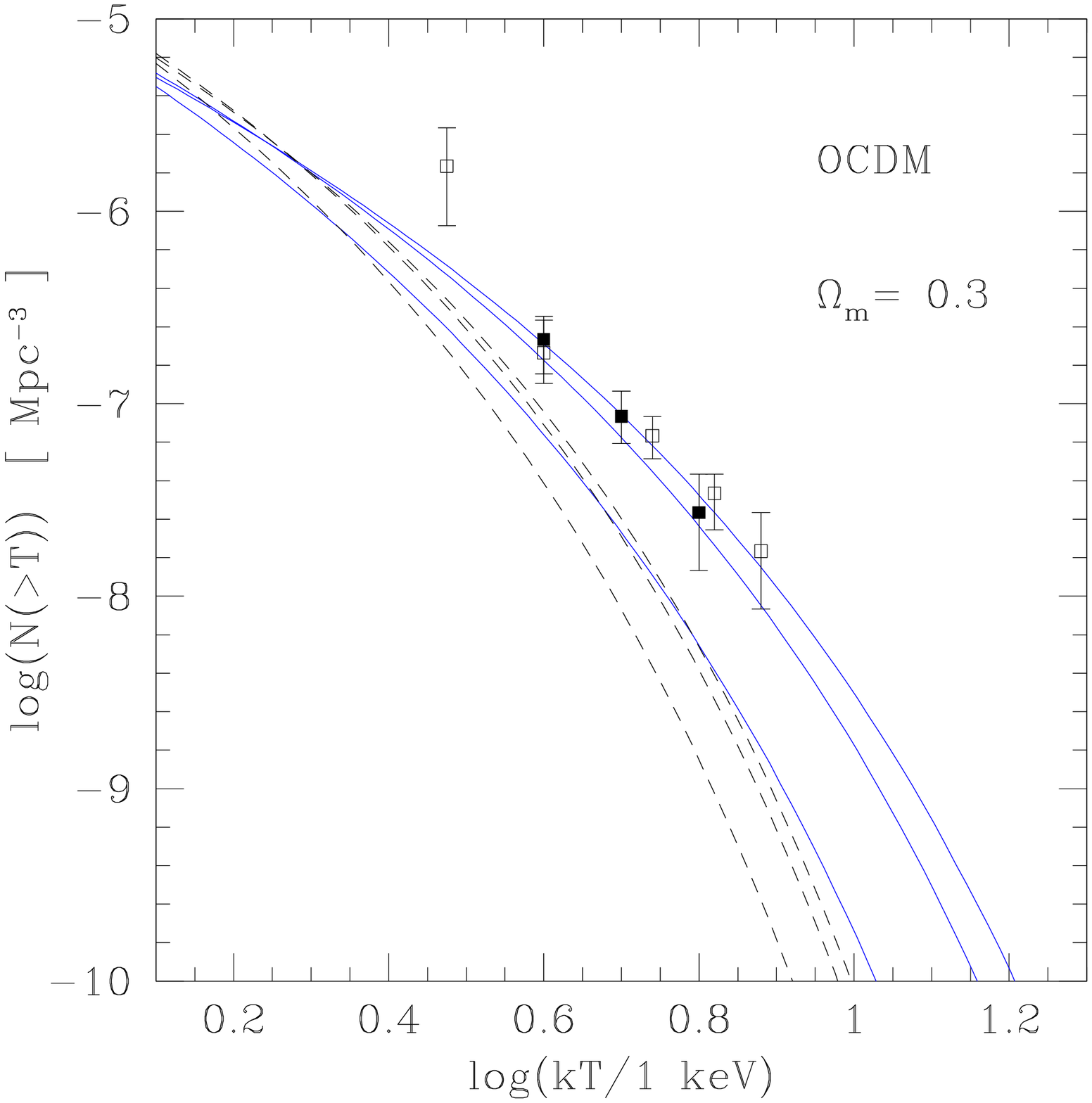}}

\caption{The comoving differential and cumulative temperature functions as in 
Fig.\ref{figTO1}, for the case of an open universe with $\Om=0.3$.}
\label{figTO03}

\end{figure}

Using $\eta(T) dT/T = \eta(M) dM/M$ we can get the cluster temperature function 
from (\ref{MT}) and (\ref{etaPS}) or (\ref{etah}). Its evolution with redshift 
is shown in Fig.\ref{figTO1} and Fig.\ref{figTO03}. Note that the temperature 
$T$ we consider in this section is the virial temperature. Indeed, for the hot 
clusters ($T \ga 1$ keV) we study here it is also the temperature of the gas 
which is heated by shocks during the gravitational collapse. In contrast, in 
cool groups ($T \la 1$ keV) the gas is also influenced by a possible preheating 
of the IGM (e.g., Valageas \& Silk 1999b) which leads to a smoother baryonic 
density profile and a larger gas temperature. This is discussed in 
Sect.\ref{Evolution of the X-ray luminosity function}. 

First, we can check that the difference between the scaling prescription and the 
PS prediction is similar to the trend we obtained for the mass functions in 
Sect.\ref{Evolution of the mass function}. In particular, we can note that 
Governato et al. (1999) found that for a standard SCDM model ($\Om=1$) normalized 
to $\sigma_8=0.5$ (which is our case) the PS prescription underestimates the 
number of hot clusters $kT>7$ keV by almost a factor 10 at $z=1$, while it 
overestimates the number of small halos. These authors also found that the 
deficiency of massive halos predicted by the PS approach gets more severe for 
smaller $\sigma_8$ (hence also at higher redshifts). In our view (see the 
discussion in VS), this is simply due to the fact that in this case one looks at 
rarer objects, farther in the cutoff of the mass function, which increases the 
discrepancy between both theoretical mass functions which have different 
exponential tails. However, the discrepancy between the PS and the scaling 
predictions is again slightly larger than what these numerical results imply at 
the large mass end.

\begin{figure}[htb]

\centerline{\epsfxsize=8 cm \epsfysize=5.5 cm \epsfbox{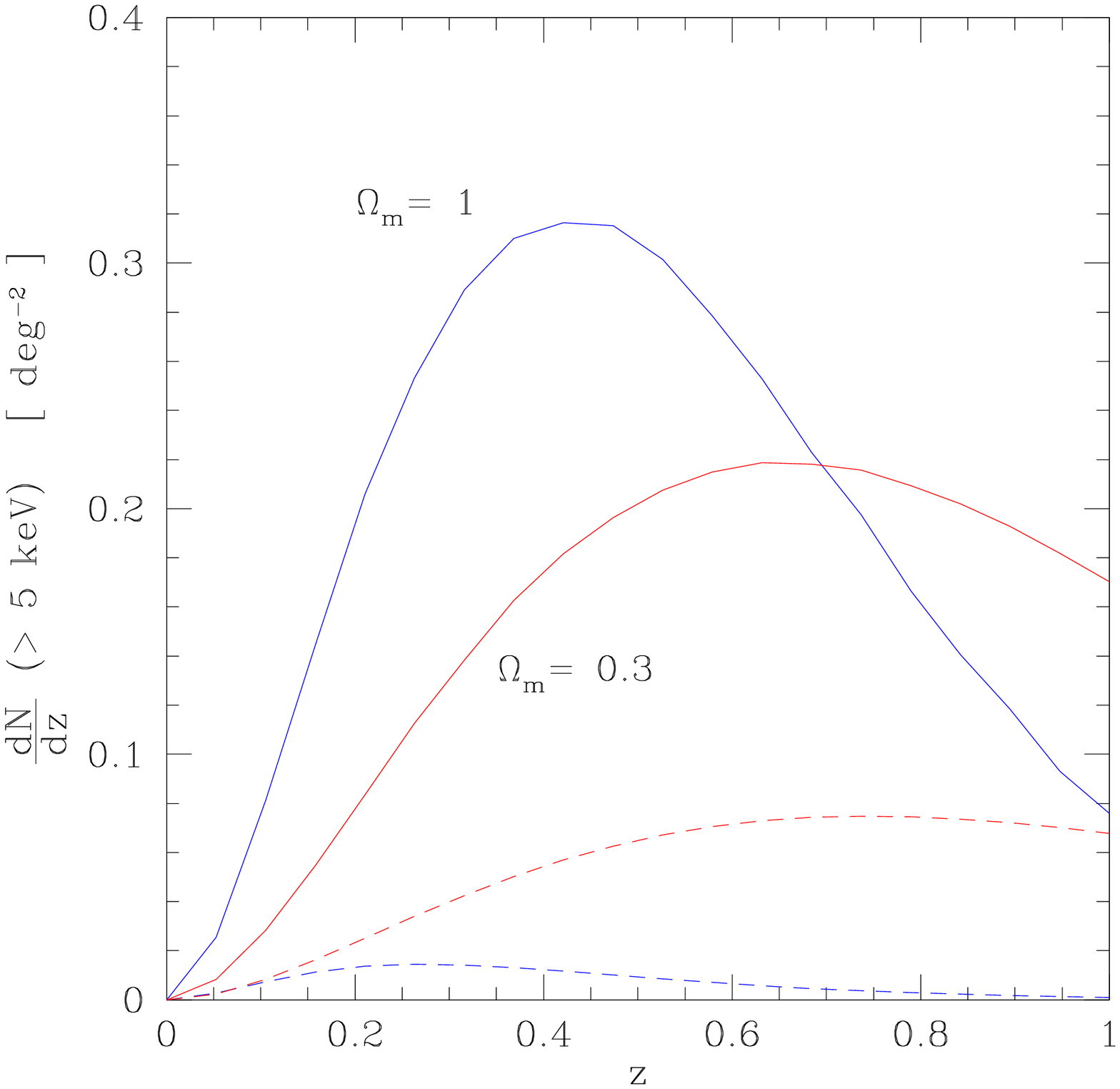}}

\caption{The redshift distribution per square degree of clusters hotter than 5 
keV, for $\Om=1$ and $\Om=0.3, \; \Ol=0$, with a CDM power-spectrum. The graph 
shows the number of clusters $d{\cal N}/dz(>5$ keV) per unit redshift interval 
and square degree. The solid lines correspond to the non-linear prescription 
while the dashed lines represent the PS formulation. The open universe 
corresponds to the slowest redshift evolution.}
\label{figNsupTz}

\end{figure}

Next, we can check that the redshift evolution of the temperature function is 
faster for the critical density universe than for the open case, in agreement 
with Fig.\ref{figM} for the mass function (see also Oukbir \& Blanchard 1997). 
Thus, the very small decline with $z$ of the observed cluster temperature 
function (Henry 1997) from $z=0.05$ to $z=0.33$ favors the open case (or more 
generally a low-density universe). However, the redshift evolution of the 
cluster temperature function we obtain in the SCDM case, is not much faster than 
the observed decline and our study shows it cannot be ruled out.
 Note that we could obtain a better agreement with the data for 
$\Om=1$ by choosing a slightly lower $\sigma_8$. However, the normalization of 
the power-spectrum we use is constrained by our previous studies of galaxies, 
quasars and Lyman-$\alpha$ clouds since we want to build a unified consistent 
model. Hence we must choose a value which provides good results for all these 
objects. Moreover, as explained in Sect.\ref{Multiplicity functions} massive 
clusters correspond to mildly non-linear scales close to the theoretical limit 
of validity of the scaling model, so that we may slightly overestimate the 
number of very massive clusters. Thus the value $\sigma_8=0.5$ used for $\Om=1$ 
(which is also the result obtained by Governato et al. 1999) seems satisfactory.

The redshift evolution of the temperature function is slower than the change in 
the mass function which was presented in Fig.\ref{figM}. This is due to the 
temperature-mass relation (\ref{MT}) which implies that $T \propto M^{2/3} \; 
\Delta_c(z)^{1/3} \; (1+z)$. Thus, the temperature which corresponds to a given 
mass increases with $z$, which enhances the redshift evolution of the mass 
function as compared with the temperature function.

The redshift distribution of clusters hotter than 5 keV, per square degree, is 
shown in Fig.\ref{figNsupTz}. Of course, our results are similar to 
Fig.\ref{figNsupMz}.

\section{Sunyaev-Zel'dovich effect}
\label{Sunyaev-Zel'dovich effect}

An indirect method to get observational constraints on the cluster temperature 
function is to measure the Sunyaev-Zel'dovich (SZ) effect (Sunyaev \& Zel'dovich 
1972) due to the hot gas in the intra-cluster medium. As compared to the X-ray 
luminosity function which we discuss below in Sect.\ref{Evolution of the X-ray 
luminosity function}, the SZ effect presents two strong advantages: it does not 
depend on the detailed density profile of the gas distribution within clusters 
and it is more sensitive to high-redshift objects. Hence in this section we 
describe our predictions for this indirect measure of the cluster multiplicity 
function. 

First, we recall that the variation of the CMB brightness at the frequency $\nu$ 
along the line-of-sight due to the SZ effect can be written as (e.g., Barbosa et 
al. 1996):
\beq
i_{\nu} = y \; j_{\nu}(\x ) ,
\eeq
where $j_{\nu}$ describes the spectral form of the distortion, independent of 
the cluster, and $y$ is the Compton parameter given by an integration along the 
line-of-sight through the cluster:
\beq
y = \int n_e \sigma_T \frac{kT_g}{m_e c^2} dl .
\label{y1}
\eeq
Here, $T_g$ is the temperature of the electrons in the intracluster gas (which 
we approximate by the virial temperature $T$), $m_e$ the electron mass, $n_e$ 
the electron number density and $\sigma_T=6.65 \; 10^{-25} \;\mbox{cm}^2$ the 
Thompson cross section. Defining the dimensionless frequency $\x =h_p 
\nu/kT_0=\lambda_0/\lambda$ where $h_p$ is Planck constant and $\lambda_0=5.28$ 
mm for $T_0=2.726$ K, $T_0$ being the present temperature of the CMB, one can 
write:
\beq
j_{\nu}(\x ) = 2 \frac{(kT_0)^3}{(h_p c)^2} \frac{\x ^4 e^\x }{(e^\x -1)^2}
 \left( \frac{\x }{\tanh(\x /2)}-4 \right) .
\eeq
The flux $S_{\nu}(\x )$ of the cluster, in mJy $=10^{-26} \; \mbox{erg s}^{-1} 
\mbox{cm}^{-2} \mbox{Hz}^{-1}$, is simply the integral of $i_{\nu}$ over the 
solid angle subtended by the cluster:
\beq
S_{\nu}(\x ) = j_{\nu}(\x ) \; r_d(z)^{-2} \; \int n_e \sigma_T \frac{kT_g}{m_e c^2} 
dV    
\label{Snu}
\eeq
where:
\[
r_d(z) = \frac{2c}{H_0 \Om^2 (1+z)^2} \left[\Om z+(\Om-2) \left(\sqrt{1+\Om z}-1 
\right) \right]
\]
is the angular distance of the cluster, located at redshift $z$. Hence, the 
total flux observed from an unresolved cluster depends only on the mass of gas 
at the temperature $T_g$, and not on the density profile. Moreover, we can see 
that $y$ only depends on the physical properties of the cluster, and not on its 
redshift. Hence it is very sensitive to the cluster populations at high 
redshifts, which contribute in the same manner as close clusters, which means it 
is a useful tool to study the distant universe. Finally, in the long wavelength 
regime ($\x \rightarrow 0$) the negative fluctuation of the CMB spectrum is 
simply given by:
\beq
\frac{\Delta T}{T} = -2 y .
\eeq

\begin{figure}[htb]

\centerline{\epsfxsize=8 cm \epsfysize=5.5 cm \epsfbox{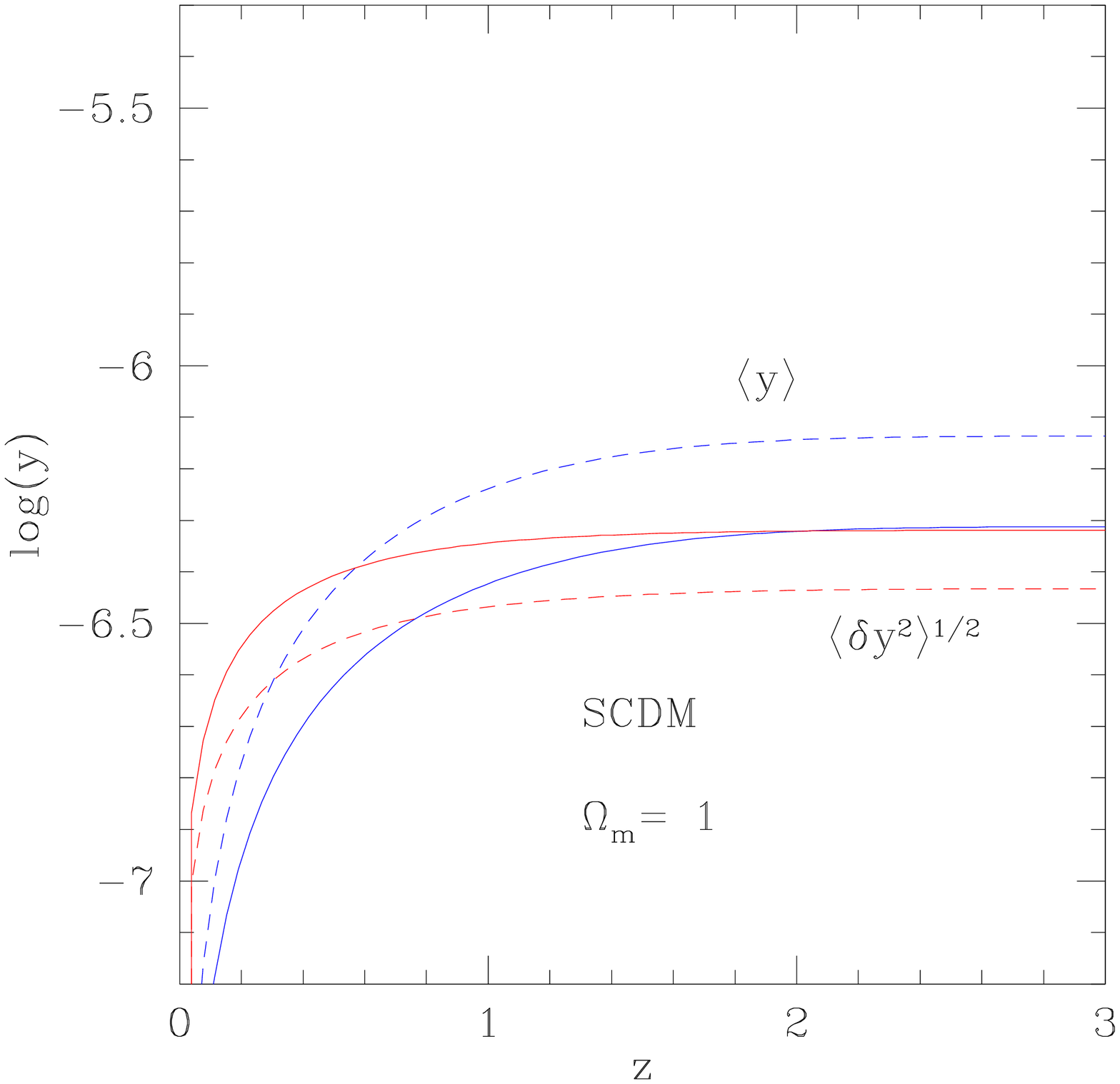}}
\centerline{\epsfxsize=8 cm \epsfysize=5.5 cm \epsfbox{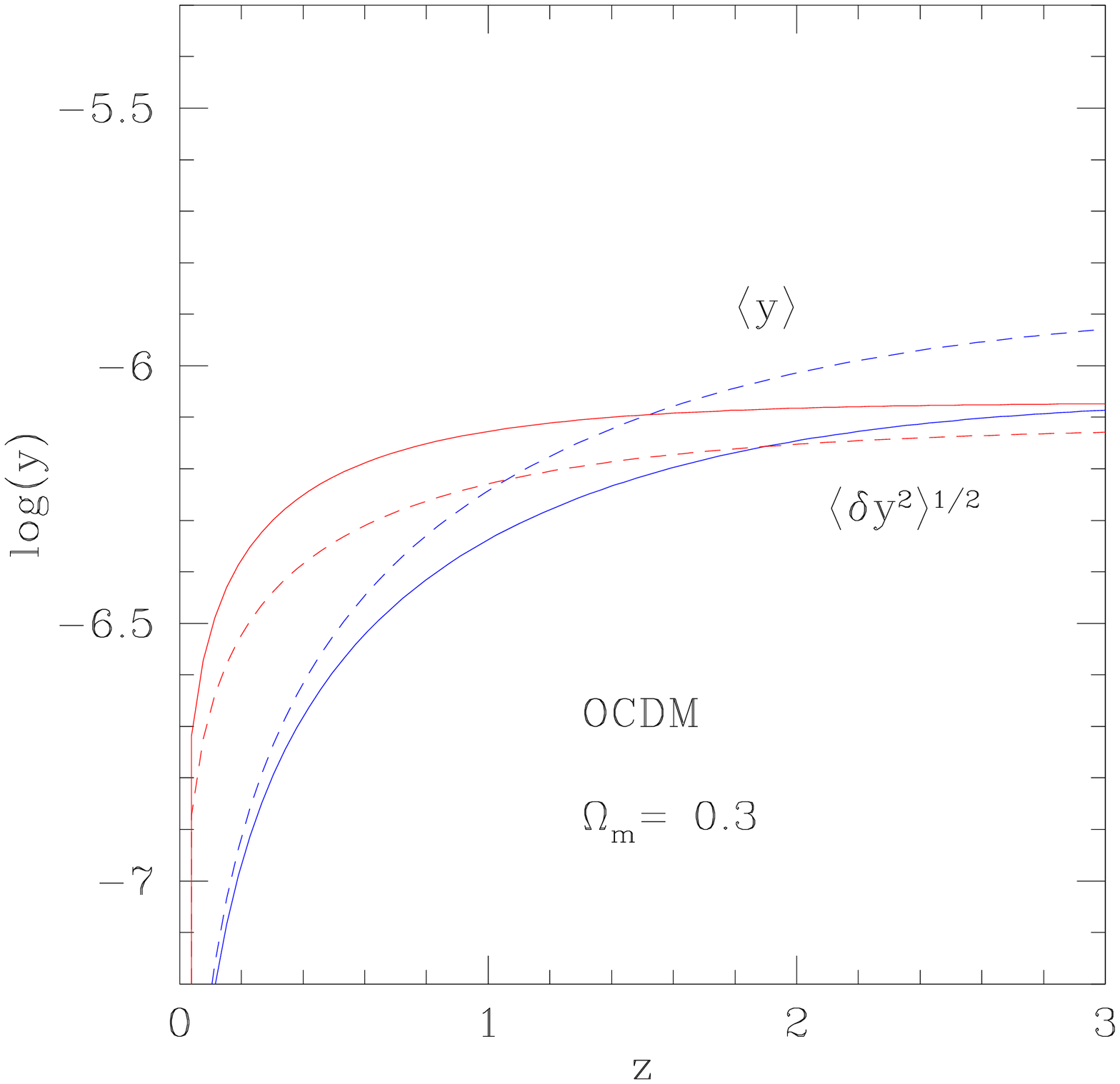}}

\caption{{\it Upper panel}: the Compton parameter $y$, and its fluctuations 
$\lag \delta y^2 \rag^{1/2}$, on a line-of-sight from $z=0$ up to the redshift 
$z$, for the case $\Om=1$. The solid lines correspond to the scaling formulation 
and the dashed lines to the PS prescription. In both cases the fluctuation $\lag 
\delta y^2 \rag^{1/2}$ is the curve which shows the steepest rise at $z=0$. {\it 
Lower panel}: same curves for an open universe with $\Om=0.3$.}
\label{figySZ}

\end{figure}

\begin{figure}[htb]

\centerline{\epsfxsize=8 cm \epsfysize=5.5 cm \epsfbox{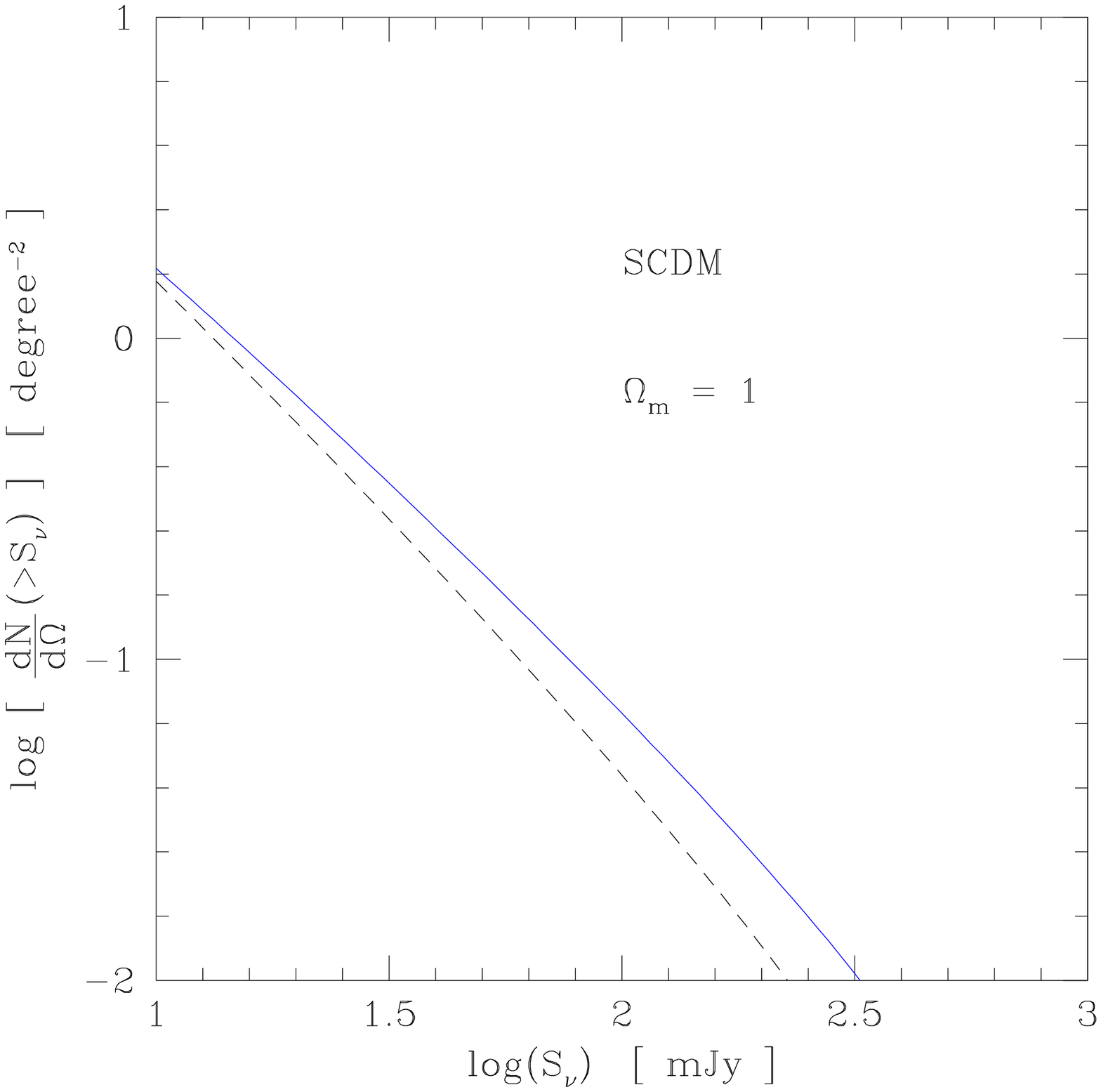}}
\centerline{\epsfxsize=8 cm \epsfysize=5.5 cm \epsfbox{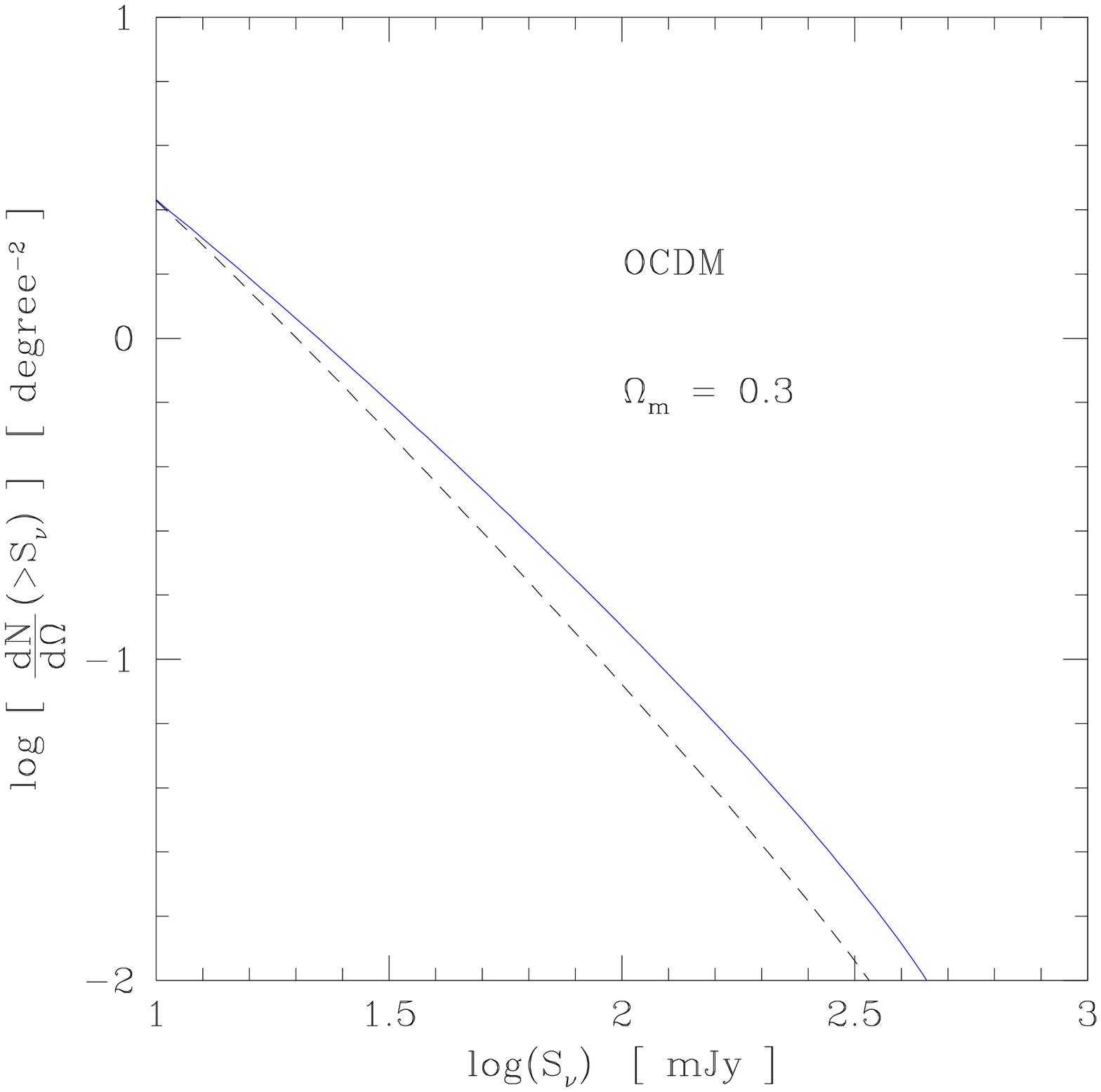}}

\caption{{\it Upper panel}: the SZ source counts at $\lambda=0.75$ mm for the 
SCDM case. The solid line corresponds to the scaling formulation and the dashed 
line to the PS prescription. {\it Lower panel}: same curves for an open 
universe.}
\label{figSZnu}

\end{figure}

The total mean Compton parameter $\lag y \rag$, 
averaged over all lines of sight,
which only depends on the 
temperature distribution of the gas, can be obtained from (\ref{y1}) :
\beq
\lag y \rag = \int c dt \int \sigma_T \frac{kT_g}{m_e c^2} 
\frac{d\overline{n}_e}{dT_g} dT_g
\label{y2}
\eeq
where $\overline{n}_e$ is the mean electron number density. Whence:
\beq
\begin{array}{l}  
{\displaystyle \lag y \rag = \int_0^{\infty} dz \frac{dt}{dz} c \sigma_T 
(1+z)^3} \\ \\ {\displaystyle  \hspace{2cm} \times \int_{M_i}^{\infty} 
\frac{dM}{M} \eta(M) \frac{\Ob}{\Om} \frac{M}{\mu_1 m_p} \frac{kT_g}{m_e c^2} .}
\end{array}
\label{y3}
\eeq
The factor $(1+z)^3$ transforms the comoving mass function $\eta(M)$ we have used so 
far, into the mass function in proper coordinates. The cutoff at $M_i(z)$ 
(defined by $t_{cool} < t_H$) is due to the fact that in small halos the gas 
cools in less than one Hubble time $t_H$. The mass $M_i(z)$ we get is typically 
of the order of $10^{12}\;M_{\odot}$.

An alternative way to get $\lag y \rag$ is to divide the line-of-sight into 
small length elements $\Delta l_i$, so that:
\beq
\lag y \rag = \lag \sum_i \hat{y}_i \rag   \;\;\; \mbox{and} \;\;\; \lag y^2 
\rag = \lag \left( \sum_i \hat{y}_i \right)^2 \rag
\label{y4} 
\eeq
where $\hat{y}_i$ is the Compton distortion due to the line element $\Delta l_i$ in 
one random realization. In this way, we recover the previous formula (\ref{y3}) 
for $\lag y \rag$ and we get the fluctuations of $y$:
\beq
\lag \delta y^2 \rag = \lag (y-\lag y \rag)^2 \rag
\label{y5}
\eeq
with:
\beq
\begin{array}{l}
{\displaystyle \lag \delta y^2 \rag = \int_0^{\infty} dz \frac{dt}{dz} c \sigma_T (1+z)^3 }  \\ \\ {\displaystyle \hspace{0.7cm} \times \Biggl \lbrace \int_{M_i}^{\infty} \frac{dM}{M} \eta(M) 
\frac{\sigma_T}{\pi R^2} \left( \frac{\Ob}{\Om} 
\frac{M}{\mu_1 m_p} \frac{kT_g}{m_e c^2} \right)^2 } \\ \\ {\displaystyle \hspace{1.3cm} + \sigma_T \xi_{cc}(R_{cl},z) R_{cl}(z) (1+z)^3 } \\ \\ {\displaystyle \hspace{2.3cm} \times \left[ \int_{M_i}^{\infty} \frac{dM}{M} \eta(M) \frac{\Ob}{\Om} 
\frac{M}{\mu_1 m_p} \frac{kT_g}{m_e c^2} \right]^2 \Biggl \rbrace }
\label{y6}
\end{array}
\eeq
The first term is due to the Poisson fluctuations of the number and the mass of 
clusters in each line element while the second term arises from the 
correlations $\xi_{cc}(r,z)$ among clusters, and $R_{cl}(z)$ is the typical radius of clusters. The first term scales as the number density $n_{cl}$ of clusters while the second term scales as $n_{cl} (\xi_{cc}(R) n_{cl} R_{cl}^3)$. Thus, the Poisson fluctuations are greater since the number density of clusters is small ($\xi_{cc}(R) n_{cl} R_{cl}^3 \ll 1$). We show in Fig.\ref{figySZ} the mean 
Compton parameter $\lag y \rag$ and its fluctuations $\lag \delta y^2 
\rag^{1/2}$ on a line-of-sight from $z=0$ up to the redshift $z$. The Compton 
parameter $\lag y \rag$ is larger for the open universe because the cluster mass 
function declines more slowly at higher $z$ and the line-element is slightly 
larger. Since the number of clusters on the line-of-sight is rather small (see 
Fig.\ref{figNls}) the fluctuations of $y$ are of the same magnitude as the mean 
$\lag y \rag$. Of course $\lag \delta y^2 \rag^{1/2}$ is much larger than $\lag 
y \rag$ for $z \simeq 0$ since the number of clusters tends to 0 in this limit. 
In particular, at $z \simeq 0$ we have:
\beq
\lag y \rag \propto \left[ (1+z)^{3/2} -1 \right] \hspace{0.4cm} \mbox{and} 
\hspace{0.4cm} \lag \delta y^2 \rag^{1/2} \propto \lag y \rag^{1/2} .
\eeq
Note that the COBE/FIRAS upper limit is $\lag y \rag < 1.5 \times 10^{-5}$ 
(Fixsen et al. 1996). Thus, in both cosmologies the Sunyaev-Zel'dovich effect as 
measured by $y$ is still one order of magnitude below the upper limit provided 
by present observations. Such a distorsion, however, will be within the reach of 
the  MAP and PLANCK projects.

\begin{figure}[htb]

\centerline{\epsfxsize=8 cm \epsfysize=5.5 cm \epsfbox{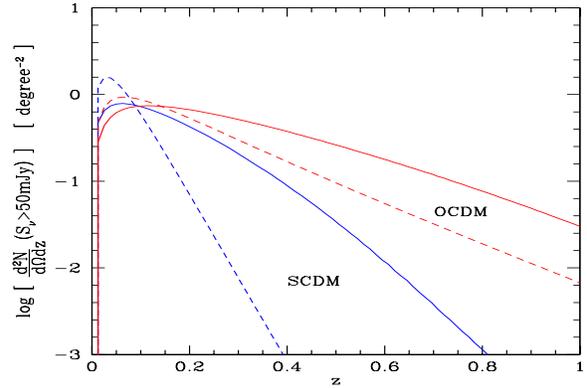}}

\caption{The redshift distribution of SZ source counts for the SCDM and OCDM 
scenarios at $\lambda=0.75$ mm. The solid lines are the scaling formulation and 
the dashed lines the PS prescription. The low-density cosmology corresponds to 
the slowest decline at large $z$ of the SZ counts.}
\label{figSZnuz}

\end{figure}

From the cluster mass function we can derive the flux density distribution of 
clusters $\eta(S_{\nu})dS_{\nu}/S_{\nu}$. Hence, the number of SZ sources per 
unit solid angle on the sky with a total monochromatic flux greater than 
$S_{\nu}$ (we assume that the objects are unresolved) is
\beq
\frac{d{\cal N}}{d\Omega} (>S_{\nu}) = \int dz \frac{dV}{d\Omega dz} 
\int_{M_i'}^{\infty} \eta(M,z) \frac{dM}{M} .
\eeq
The cutoff $M_i'(S_{\nu},z)$ corresponds to the threshold $S_{\nu}$. The SZ 
source counts at $\lambda=0.75$ mm are shown in Fig.\ref{figSZnu} as a function 
a $S_{\nu}$ for both cosmologies. As for the Compton parameter $y$, the open 
universe leads to slightly higher counts because of the slower decline of the 
cluster multiplicity function and of the larger volume element.  

The redshift distribution of these source counts is displayed in 
Fig.\ref{figSZnuz}. Of course, as explained above, the open universe predicts a 
slower decline at large $z$ of the source counts than for a SCDM scenario.

\section{Evolution of the X-ray luminosity function}
\label{Evolution of the X-ray luminosity function}

Although observations can provide an estimate of the cluster temperature 
function they can more easily give the X-ray luminosity function, since it is 
easier to measure the luminosity of a distant cluster than its temperature. 
Hence we describe in this section a model to obtain the X-ray luminosity of 
clusters of galaxies. Moreover, the temperature-luminosity relation also 
contains some interesting information on a possible reheating of the IGM.

\subsection{The temperature-luminosity relation}
\label{The temperature-luminosity relation}

\subsubsection{Breaking the simple scale-invariance}
\label{Breaking the simple scale-invariance}

The bolometric X-ray luminosity $L_{bol}$ of a cluster of volume $V$ is
\beq
L_{bol} = \epsilon \int_V n_e^2 \Lambda_b(T_g) dV
\label{Lbol1}
\eeq
where $\epsilon$ is a constant of order unity, 
$n_e$ is the electron number density and $\Lambda_b(T_g)$ is the 
emissivity function (in erg cm$^3$ s$^{-1}$) for a gas at the temperature $T_g$. 
Thus, contrary to the Sunyaev-Zel'dovich effect, see (\ref{Snu}), the X-ray 
luminosity strongly depends on the density profile of the hot gas within the 
cluster. Using (\ref{MT}) and the fact that $\Lambda_b(T) \propto \sqrt{T}$ for 
bremsstrahlung one expects (Kaiser 1986):
\beq
L_{bol} \propto T_g^2 .
\label{Lbol2}
\eeq
Note that this power-law behaviour is related to the scale-invariance of 
clusters: more massive objects are identical to smaller ones after a simple 
rescaling. However, observations show a much steeper slope $L_{bol} \propto 
T_g^{2.88}$ (Arnaud \& Evrard 1999) and suggest a bend in the 
temperature-luminosity relation. Hence some physics is missing in the derivation 
leading to (\ref{Lbol2}). In particular, one needs to break the scaling laws 
which led to (\ref{Lbol2}) through the introduction of additional dimensional 
quantities. It has been suggested in the litterature (e.g., Evrard \& Henry 
1991; Cavaliere et al. 1997; Ponman et al. 1999; Lloyd-Davies et al. 2000) that one 
needs to take into account the reheating of the IGM, prior to cluster formation, 
which raises its entropy and can modify the gas dynamics in cool clusters. 
Indeed, if this entropy ``floor'' is sufficiently large, the gas is heated by the 
adiabatic compression up to a temperature which can be as large as the virial 
temperature of the halo. In this case the density profile of the gas is much 
smoother than the distribution of the dark matter, which diminishes the 
luminosity of these cool clusters and modifies the relation $T_g - L_{bol}$. 
Such a reheating of the IGM by quasars or supernovae was studied in Valageas \& 
Silk (1999b) where it was shown that the energy provided by quasars may be 
sufficient to reheat the IGM. The energy delivered by supernovae was found by 
the latter authors to be rather small, but a more thorough estimate is 
under way. In this article, we simply assume that such processes have reheated 
the gas to a characteristic temperature $T_{ad} \sim 0.4$ keV. This breaks the 
simple scaling laws and can lead to a non-trivial temperature-luminosity 
relation.

\subsubsection{Density profiles of the gas and of the dark matter}
\label{Density profiles of the gas and of the dark matter}

In order to obtain the distribution of the gas, we need the density profile of 
the dark matter, which is gravitationally dominant. Here we assume that the 
halos obey the density profile obtained in high-resolution simulations by Moore 
et al. (1999b):
\beq
\rho(r) = \frac{\rho_c}{(r/r_s)^{1.5} \left[ 1 + (r/r_s)^{1.5} \right]} 
\hspace{0.4cm} \mbox{with} \hspace{0.4cm} r_s=\frac{R}{c}
\label{Moore1}
\eeq
where $c$ is the concentration parameter. We use $c=4$ as in Moore et al. (1999b), although this parameter may slightly depend on the mass of the cluster. Moreover, we assume hydrostatic equilibrium for the gas. We consider two models for the temperature profile $T_g(r)$ of the gas.

First, we write:
\beq
T_g(r)= T_s(r) + T_{ad} \left( \frac{\rho_g(r)}{(1+\Delta_c) \rhoa_b} 
\right)^{\gam_s-1}
\label{Tgas1}
\eeq
where $\gam_s=5/3$. The first term on the right hand side in (\ref{Tgas1}) describes 
non-adiabatic gravitational heating through shocks during the formation of the 
cluster. The second term takes into account the reheating of the gas, before the 
formation of the cluster, and its subsequent heating through the adiabatic 
compression during the infall. Note that the relation (\ref{Tgas1}) assumes that 
the thermal conduction is small (e.g., Sarazin 1988) so that the gas is not 
exactly isothermal. In a fashion similar to (\ref{Tvir}) we write:
\beq
T_s(r) = \frac{{\cal G} \mu m_p M(<r)}{2 k r} = T \; \frac{ \ln \left( 
1+(r/r_s)^{1.5} \right) }{ \ln \left( 1+c^{1.5} \right) } \; \frac{R}{r}
\label{Tsr}
\eeq
which also measures the depth of the potential well. This also satisfies the 
Jeans eq.(\ref{Jeans}) within a factor 2. Then, the distribution of the gas is 
given by the condition of hydrostatic equilibrium:
\beq
\frac{dP}{dr} = - \; \rho_g \; \frac{{\cal G} M(<r)}{r^2} \hspace{0.5cm} 
\mbox{with} \hspace{0.5cm} P = \frac{\rho_g k T_g}{\mu m_p} .
\label{pressure}
\eeq
However, in practice we use a simplified model. From (\ref{pressure}) one can 
check that at large radii where $T_g \simeq T_s(r)$ the gas follows the dark 
matter while at small radii below $R_{core}$ where both terms in (\ref{Tgas1}) 
are equal, the gas density saturates. Indeed, when $T_g \gg T_s$ the gas no longer falls into the potential well and a core with a nearly constant baryon 
density appears. This model is similar to those used in Cavaliere et 
al. (1997,1999) or Wu et al. (1999) except that these authors use the dark matter 
density profiles given by Navarro et al. (1996) and they do not take into account 
the first term on the right hand side in (\ref{Tgas1}) (they use a 
polytropic equation of state). However, in all cases the behaviour of the gas 
distribution is the same: for smaller halos the preheating becomes more 
important and the gas density profile gets flatter which decreases the X-ray 
luminosity. Here, in order to simplify the calculations we approximate the gas 
distribution by the same profile (\ref{Moore1}) as 
for the dark matter in the outer 
parts $r>R_{core}$ and by a constant value in the core $r<R_{core}$:
\beq
\left\{ \begin{array}{ll} {\displaystyle r > R_{core} \; : } & {\displaystyle 
\rho_g(r) = \frac{\Ob}{\Om} \; \rho(r) } \\ \\ {\displaystyle r < R_{core} \; : 
} & {\displaystyle \rho_g(r) = \rho_{core} = \frac{\Ob}{\Om} \; 
\frac{M(<R_{core})}{4 \pi R_{core}^3/3} } \end{array} \right.
\label{core}
\eeq
which ensures that the total mass is conserved.

Second, we consider an alternative model where the temperature of the gas is 
given by:
\beq
T_g(r)= T_s(r) + T_{ad} .
\label{Tgas2}
\eeq
As in (\ref{Tgas1}) two processes govern the temperature of the gas: 
gravitational heating (first term) and a second unspecified source of energy 
(e.g., SNe or QSOs) which breaks the scaling law described in Sect.\ref{Breaking 
the simple scale-invariance} (second term). However, in contrast to 
(\ref{Tgas1}), here we take this additional term to be constant. Hence this 
model corresponds to a uniform ``post-heating'' of the gas which occurs after 
the formation of the cluster, for instance through the SNe which may eject some 
energy into the intra-cluster medium during and after the infall of the gas into 
the potential well of the cluster. From another point of view, this can also be 
seen as a specific case of the ``pre-heating'' model (\ref{Tgas1}) with a 
different equation of state: $\gam_s \simeq 1$. Then, this allows us to estimate 
the sensitivity of our model (\ref{Tgas1}) to the assumed value for $\gam_s$. 
Next, from the relation (\ref{Tgas2}) we proceed exactly as from (\ref{Tgas1}) 
to obtain the density distribution of the gas, described by a core radius 
$R_{core}$ as in (\ref{core}), which is now defined by $T_s(R_{core})=T_{ad}$.

\subsubsection{Cooling radius}
\label{Cooling radius}

\begin{figure}[htb]

\centerline{\epsfxsize=8 cm \epsfysize=5.5 cm \epsfbox{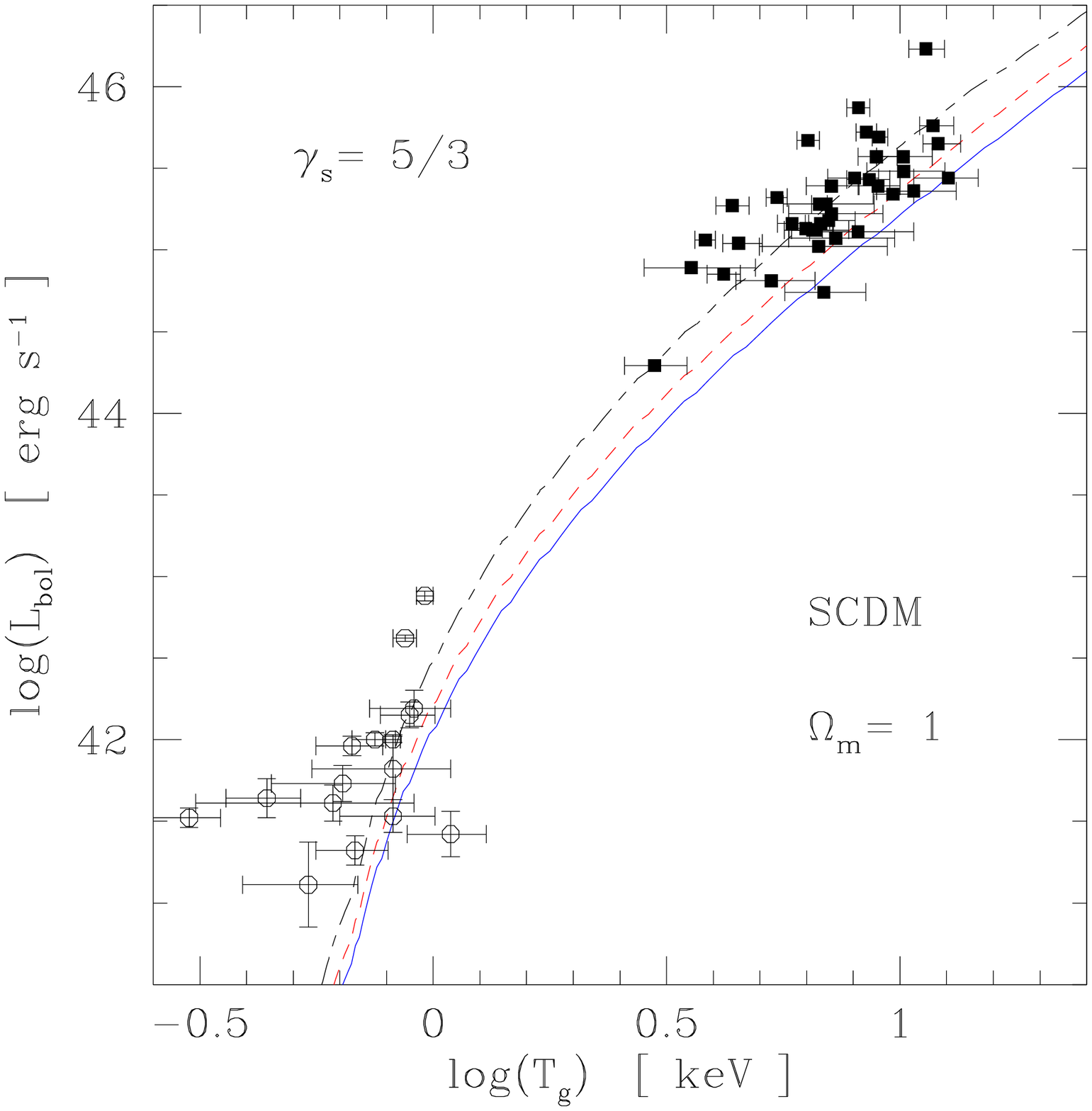}}
\centerline{\epsfxsize=8 cm \epsfysize=5.5 cm \epsfbox{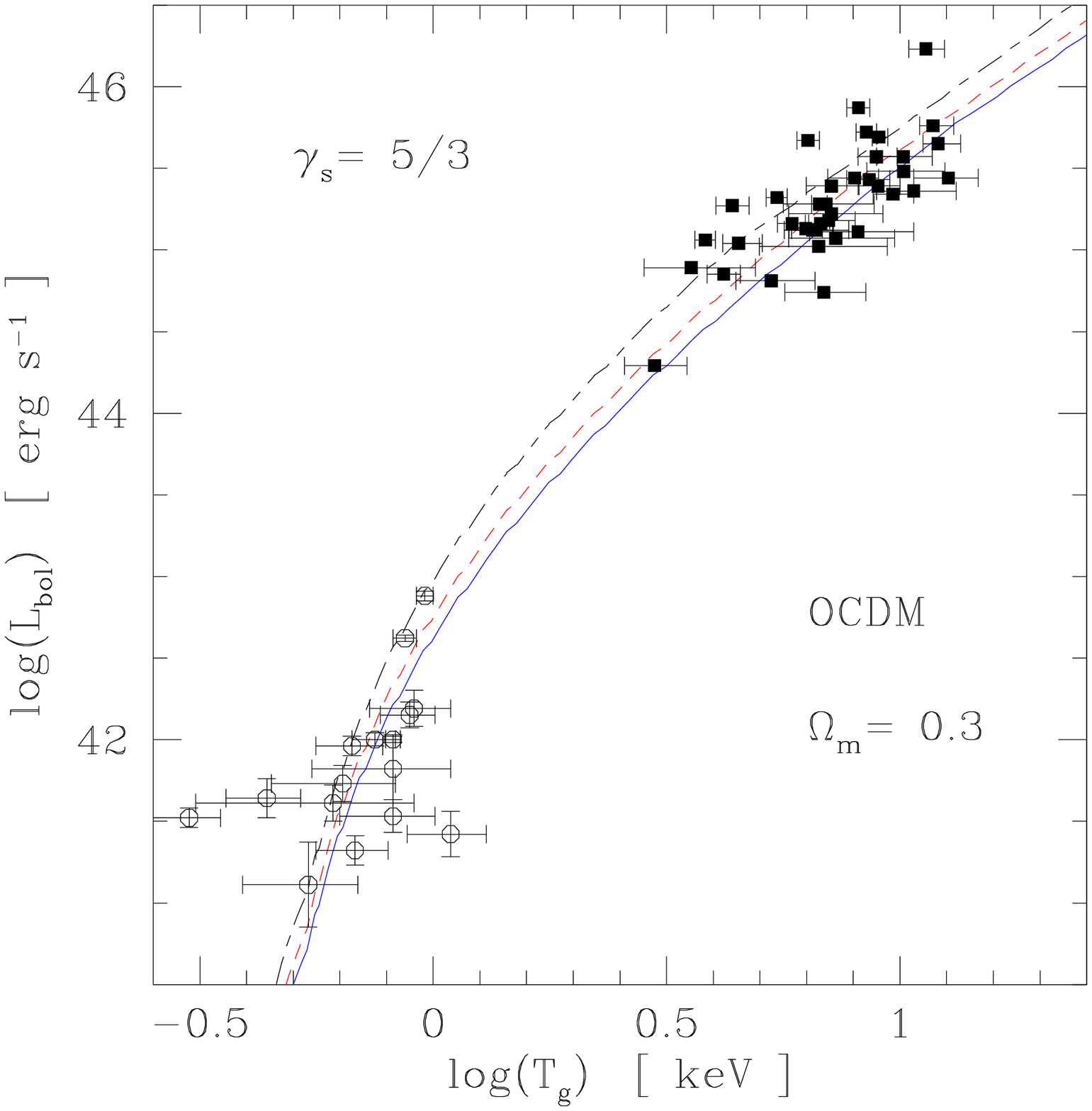}}

\caption{The relation temperature - bolometric luminosity for clusters in the 
case $\gamma_s=5/3$ for the model (\ref{Tgas1}) with $T_{ad}=0.5$ keV (resp. 
$T_{ad}=0.4$ keV) for the SCDM (resp. OCDM) scenario. The solid, dashed and 
dot-dashed lines correspond to $z=0$, $z=0.33$ and $z=1$ respectively. The data 
points are from Mushotzky \& Scharf (1997) ($0.14<z<0.55$) for clusters and from 
Ponman et al. (1996) ($z<0.05$) for groups. The upper panel corresponds to a 
critical density universe (SCDM) and the lower panel to an open universe 
(OCDM).}
\label{figTLclus53}

\end{figure}

\begin{figure}[htb]

\centerline{\epsfxsize=8 cm \epsfysize=5.5 cm \epsfbox{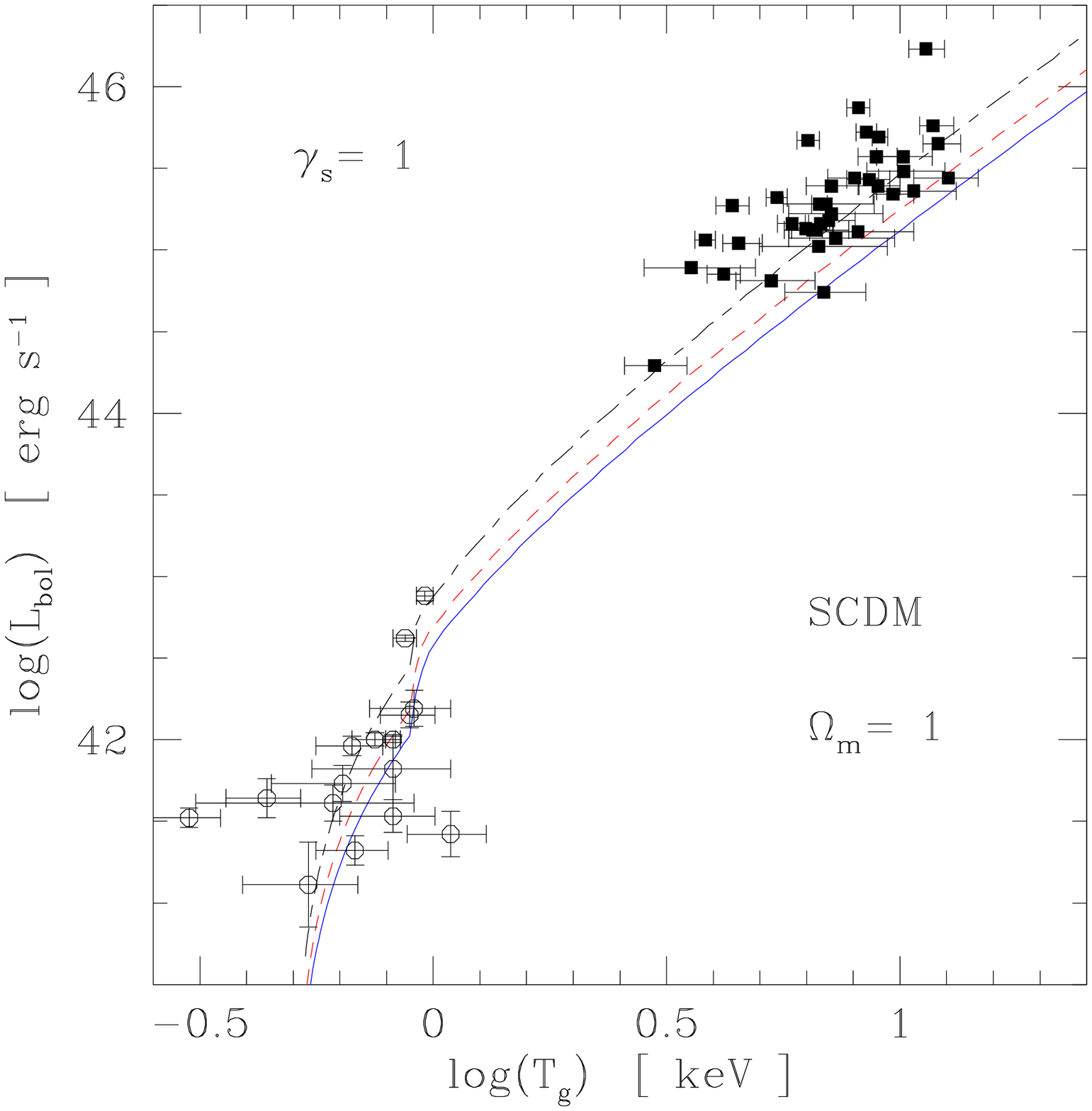}}
\centerline{\epsfxsize=8 cm \epsfysize=5.5 cm \epsfbox{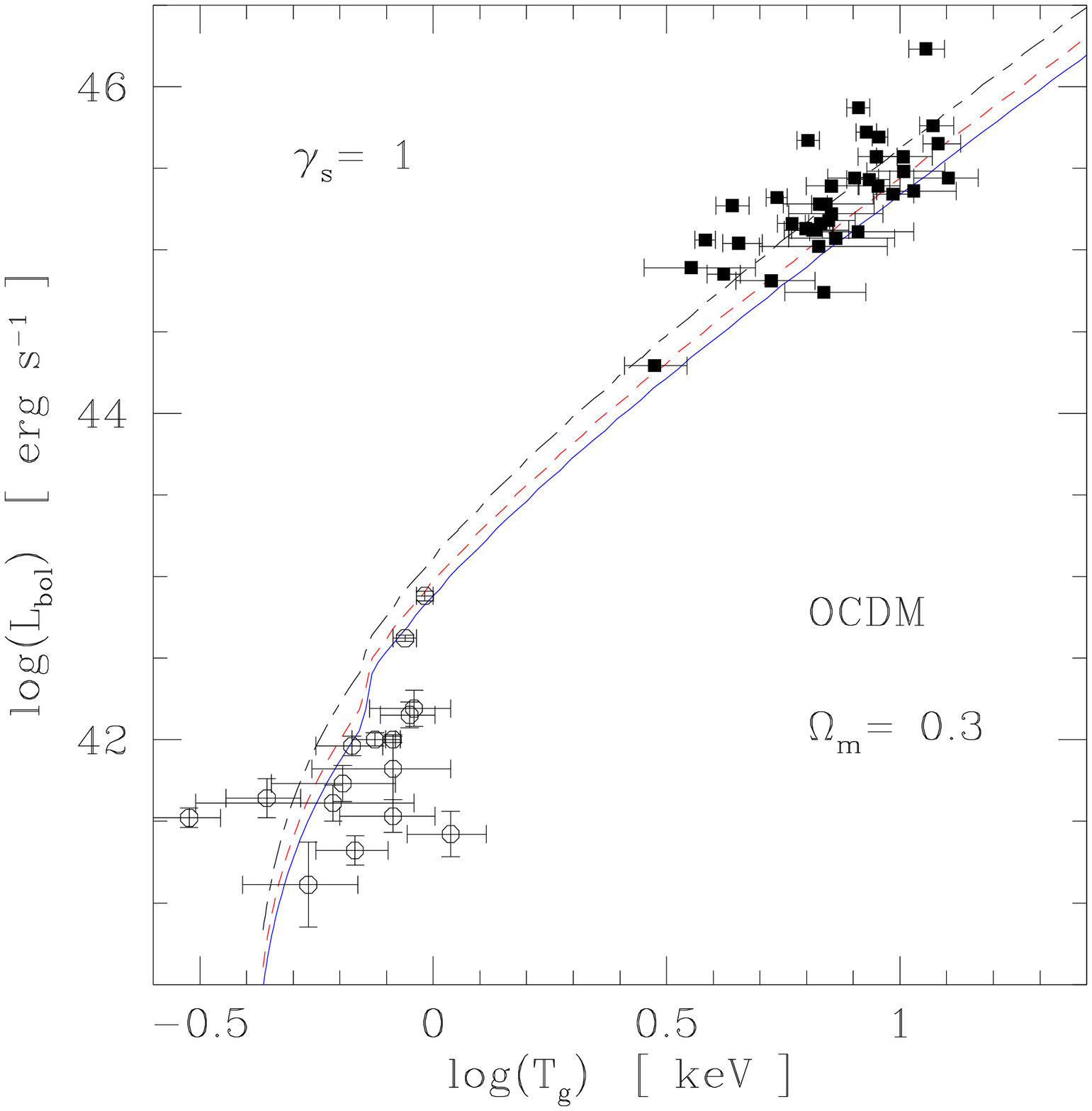}}

\caption{The relation temperature - bolometric luminosity as in 
Fig.\ref{figTLclus53} for the model (\ref{Tgas2}) (i.e. $\gamma_s=1$) with the 
same values for $T_{ad}$. A larger value for $T_{ad}$ would improve the 
agreement with observations at low $T_g$ for the OCDM scenario.}
\label{figTLclus1}

\end{figure}

In addition to the core described above which appears for the gas distribution, 
cooling may affect the distribution of hot X-ray emitting gas. Indeed, in the 
inner parts of the cluster the density can be large enough to lead to a small 
cooling time so that a cooling flow develops. Then, some of the gas forms a cold 
component which does not emit in X-ray any longer. Thus, we define the cooling 
radius $R_{cool}$ as the shell where the cooling time $t_{cool}$ becomes equal 
to the Hubble time $t_H(z)$:
\beq
t_{cool} = t_H \hspace{0.5cm} \mbox{with} \hspace{0.5cm} t_{cool} = \frac{3 
\mu_1^2 m_p k T_g}{2\mu \rho_g \Lambda_c(T_g)} ,
\label{tcool}
\eeq
where $\Lambda_c(T_g)$ is the cooling function. At large radii $r>R_{cool}$ the 
local cooling time is larger than the Hubble time. In this case, the gas distribution 
and the temperature have not had time to evolve much and the X-ray emissivity is 
proportional to $\rho_g^2 \; \Lambda_b(T_g)$, see (\ref{Lbol1}). On the other 
hand, within the cooling radius $R_{cool}$ the gas has had time to cool and to form 
dense cold clouds. However, we consider that some of the gas is still hot and 
emits in X-ray as cooling does not proceed in a uniform fashion (Nulsen 1986; 
Teyssier et al. 1997; Waxman \& Miralda-Escude 1995). Indeed, the cooling 
instability leads to a wide range of gas temperatures and densities as overdense 
regions cool faster and contract (because of the pressure of the surrounding 
gas) which increases even further their density contrast. Then, some of the gas
is simply removed from the 
X-ray emitting
component
as these dense cold clouds decouple from the hot 
phase. On the other hand, the temperature of the hot gas remains close to 
$T_s(r)$, introduced in (\ref{Tsr}), through adiabatic compression and 
possible gravitational heating, see Nulsen (1986, 1998) for detailed models of 
this multiphase ICM. The density of the hot component must be of order 
$\rho_{cool}$, defined by the condition $t_{cool}=t_H$, since the density has 
not had time to decrease further yet (note that the system is not stationary). 
As this multiphase medium is connected to the outer parts of the cluster 
$r>R_{cool}$ which also provide a reservoir of matter and energy, we assume that 
the hot low-density phase is spread all over the radius $R_{cool}$. Note that 
the time-scale for hydrostatic equilibrium is $t_p \sim r/c_s \sim t_{dyn}$ for 
gas at the temperature $T_s$ (where $c_s$ is the sound speed while $t_{dyn}$ is 
the dynamical time). Since the time-scale of the infall of the gas cannot be 
smaller than $t_{dyn}$ and is actually expected to be larger (the pressure and 
possible energy injection from SNe or gravitational interactions slow down the 
motion) there should be approximate pressure equilibrium. Hence at small radii 
$r<R_{cool}$ we write for the density of the X-ray emitting gas:
\beq
r < R_{cool} \; : \hspace{0.3cm} \rho_{gX}(r) = \rho_{cool} = \rho_g(R_{cool}) < 
\rho_g(r)
\label{cool}
\eeq
while at large radii $r>R_{cool}$ we have $\rho_{gX}(r) = \rho_g(r)$. For cool 
clusters where the non-gravitational energy term $T_{ad}$ plays an important 
role, the distribution of the gas is flatter than the dark matter and it shows a 
core of constant density $\rho_{core}$. Then, the cooling time at this core 
radius $R_{core}$ is still longer than the Hubble time so that cooling plays no 
role: the luminosity of the cluster is determined by the radius $R_{core}$ due 
to the distribution of the gas itself. On the other hand, for hot clusters the 
gas follows the dark matter density profile over a large range of radii and the 
cooling time gets smaller than the Hubble time in the ``outer region'' $r > 
R_{core}$. In this case, the distribution of the X-ray emitting gas is 
characterized by the cooling radius $R_{cool}$ and the luminosity of the cluster 
is governed by the cooling criterium. Note that $R_{cool}$ mainly plays the role 
of a cutoff for the distribution of X-ray emitting gas: we would obtain similar 
results for a model where we set $\rho_{gX}(r) =0$ for $r<R_{cool}$, as 
most of the X-ray emission comes from the regions $r \sim R_{cool}$ (thus such a 
model would simply decrease the luminosity by a numerical factor $\sim 2$ which 
can be absorbed in the normalization of $t_{cool}$ for instance).

In Valageas \& Silk (1999b) we used for illustrative purposes an isothermal 
model for the dark matter and the gas distribution, with the relation 
(\ref{Tgas2}) (interpreted as $\gam_s=1$). Thus, we had:
\beq
\rho(r) \propto r^{-2} \hspace{0.3cm} \mbox{and} \hspace{0.3cm} \rho_g(r) 
\propto \rho^{-\beta} \hspace{0.3cm} \mbox{with} \hspace{0.3cm} \beta = 
\frac{T}{T+T_{ad}} ,
\label{beta}
\eeq
where $T$ is the virial temperature of the cluster defined in (\ref{Tvir}) and 
the gas distribution was obtained from the hydrostatic equilibrium condition 
(\ref{pressure}). Here, the distribution of the gas does not show a core radius 
$R_{core}$ while the X-ray emitting gas is still characterized by a cooling 
radius $R_{cool}$. However, it is easy to check that we recover a behaviour 
similar to the case discussed above. Indeed, for hot clusters with $T>3 T_{ad}$ 
the density profile of the gas is very steep ($\beta>3/4$) so that the X-ray 
emissivity is dominated by the inner parts of the cluster - as can be seen from 
(\ref{Lbol1}) and (\ref{beta}) - hence by the cooling radius $R_{cool}$. On the 
other hand, for cool clusters with $T<3 T_{ad}$ the density profile of the gas 
is rather flat ($\beta<3/4$) which means that the luminosity of the cluster is 
governed by the outer parts $r \sim R$. In this paper, for cool clusters the 
emissivity is dominated by the regions $r \sim R_{core}$ and $R_{core} = R$ for 
low temperature clusters ($T \leq T_{ad}$). 

Thus, we see that the main characteristics of clusters do not strongly depend on 
the details of the models (density profile of the dark matter halo, index 
$\gam_s$ of the ``equation of state'') as they are mainly sensitive to $T_{ad}$.

\subsubsection{Evolution of the temperature - X-ray luminosity relation}

For both models described above we obtain the X-ray luminosity from 
(\ref{Lbol1}), where $n_e$ is given by the density of the X-ray emitting gas 
$\rho_{gX}$. We 
determine the factor
$\epsilon$ by requiring (\ref{Lbol1}) 
 to reproduce the observations 
for massive clusters ($T_g > 1$ keV). We get $\epsilon=2.2$ (resp. 
$\epsilon=0.8$) for a critical density universe (resp. an open universe). We 
show in Fig.\ref{figTLclus53} and in Fig.\ref{figTLclus1} the 
temperature-luminosity relation we obtain for both models (\ref{Tgas1}) and 
(\ref{Tgas2}), with the above value of $\epsilon$ and $T_{ad}=0.5$ keV (resp. 
0.4 keV) for the SCDM (resp. OCDM) cosmology. The gas temperature used in the 
figures is $T_g=T + T_{ad}$. We can see that we get very similar results which 
agree reasonably well with observations. Of course, we could improve the 
agreement of the model (\ref{Tgas2}) (i.e. $\gamma_s=1$) with observations in 
the OCDM cosmology by using a slightly larger value for $T_{ad}$. We also note 
that the redshift evolution we obtain is very small, which is consistent with 
observations (Mushotzky \& Scharf 1997). This suggests that the ``preheating'' 
temperature $T_{ad}$ should not evolve too strongly with redshift for $z < 1$, 
which might favour supernovae as the source of energy (due to the sharp decline 
at low $z$ of the quasar luminosity function, the characteristic temperature 
$T_{ad}$ obtained within the framework of a model where the preheating is due to 
QSOs, is expected to show a faster evolution with $z$, see Valageas \& Silk 
1999b).

Thus, in order to recover a bend in the temperature - X-ray luminosity relation 
one mainly needs to introduce a new dimensional parameter, like $T_{ad}$, as 
discussed in Sect.\ref{Breaking the simple scale-invariance}. Since in this 
article we are only interested in the total luminosity of clusters the simple 
models described in Sect.\ref{Density profiles of the gas and of the dark 
matter} are sufficient for our purpose and in the following we use the model 
(\ref{Tgas1}) (i.e. $\gamma_s=5/3$). As shown by the comparison of 
Fig.\ref{figTLclus53} and Fig.\ref{figTLclus1} our predictions for the 
luminosity function are not very sensitive to the details of our model.
 We would also obtain similar results with the isothermal model 
(\ref{beta}) as in Valageas \& Silk (1999b). On the other hand, this means that 
in order to discriminate between various models one needs precise observations 
of the density profiles of the gas and of the dark matter as well as a measure 
of the gas temperature. This will be provided by the XMM mission. However, present observations seem to favour the model (\ref{Tgas1}) with a 
``preheating'' of the gas before the formation of clusters (Lloyd-Davies et 
al. 2000).

\subsection{Luminosity function}
\label{Luminosity function}

\begin{figure}[htb]

\centerline{\epsfxsize=8 cm \epsfysize=5.5 cm \epsfbox{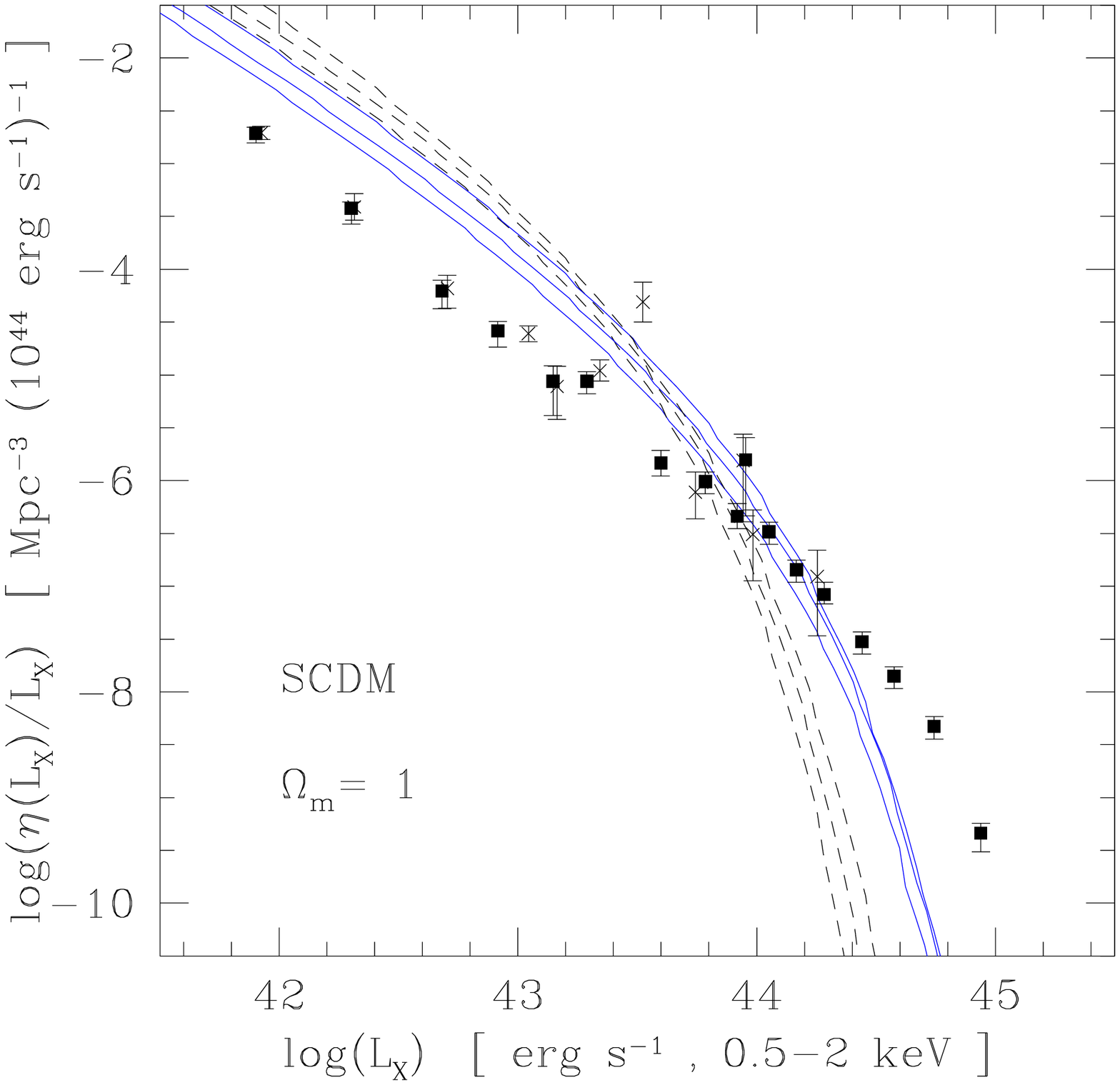}}
\centerline{\epsfxsize=8 cm \epsfysize=5.5 cm \epsfbox{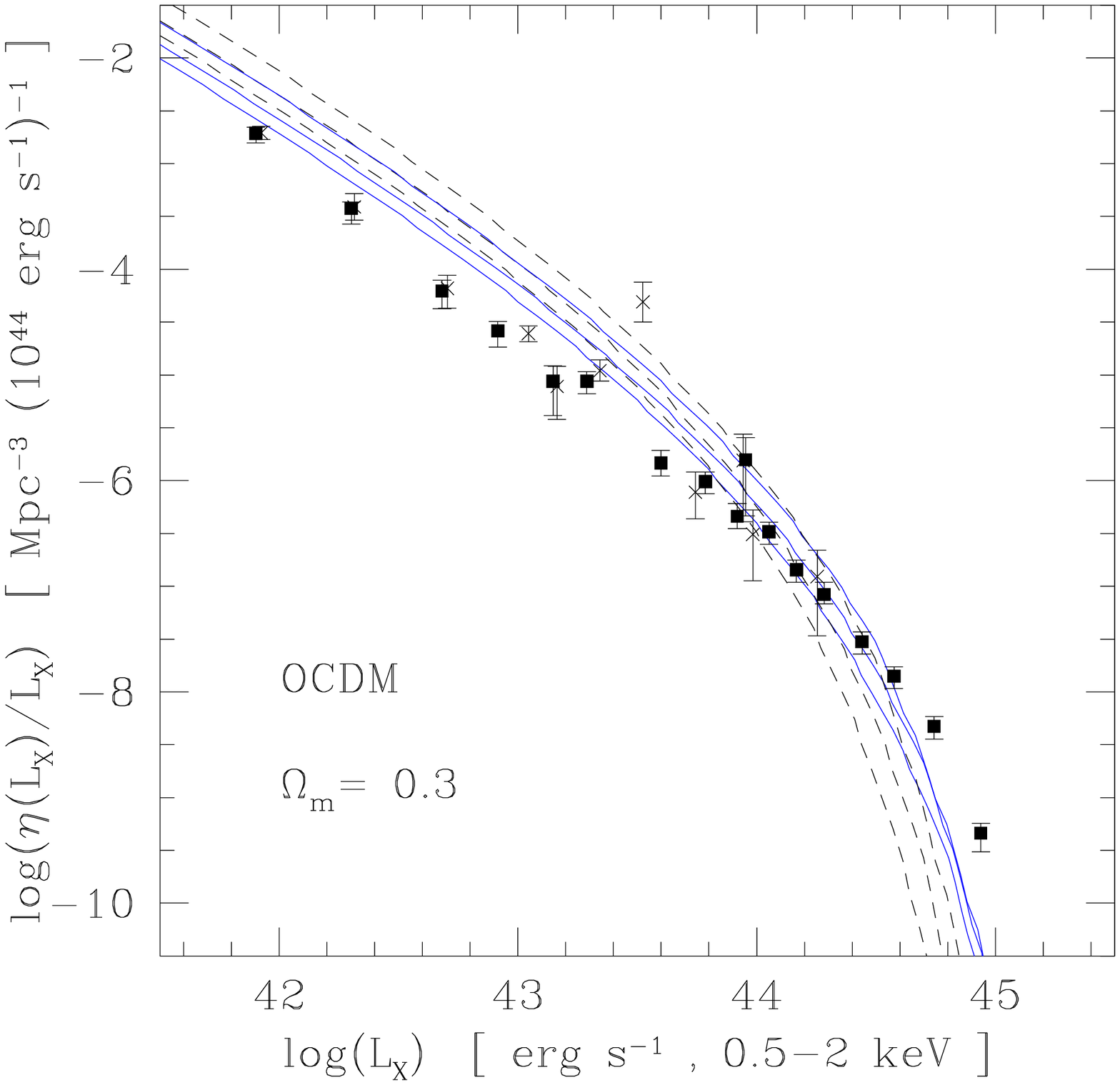}}

\caption{The comoving cluster luminosity function in the rest-frame $0.5-2$ keV 
band, at redshifts $z=0$, $z=0.33$ and $z=1$ (smaller $z$ corresponds to fewer 
faint objects). The solid lines show the scaling formulation and the dashed 
lines the PS prescription. The data points at $z=0$ are from Ebeling et 
al. (1997) (filled squares) and Burns (1996) (crosses). The upper panel 
corresponds to the SCDM scenario and the lower panel to the open universe.}
\label{figetaLclus}

\end{figure}

From the temperature-luminosity relation described in the previous section and 
the temperature multiplicity function obtained in Sect.\ref{Cluster temperature 
functions} we can derive the cluster X-ray luminosity function. Note that the 
results discussed in the following are largely independent of Sect.\ref{The 
temperature-luminosity relation} since any temperature-luminosity relation which 
agrees with observations would give similar results. From the luminosity 
$L_{bol}$ obtained in the previous section we write the luminosity $L_X$ in the 
frequency band $\nu_1 - \nu_2$ as:
\beq
L_X = L_{bol} \; \left( e^{-h_P \nu_1/kT_g} - e^{-h_P \nu_2/kT_g} \right) .
\label{LX}
\eeq
Here $h_P$ is Planck constant and we neglected the variation of the Gaunt 
factor. We compare our predictions with observations in Fig.\ref{figetaLclus} 
for both cosmological scenarios, in the rest-frame frequency band $0.5-2$ keV. 
First, we note that we recover the fact that the scaling model predicts more 
massive and bright clusters but fewer small and faint objects than the PS 
approach. Then, we see that for the critical density universe, the luminosity 
function we obtain predicts too many faint clusters. This could be cured by a 
small change of the initial power-spectrum. Of course, in a similar fashion one 
can also bring the PS prescription in agreement with observations. However, we 
prefer to keep this normalization of the power-spectrum in order to be 
consistent with our previous articles about galaxies and Lyman-$\alpha$ clouds 
and with results from numerical simulations (Governato et al. 1999). On the other 
hand, for the open universe our predictions agree reasonably well with the data. 
We can see that for both cosmological scenarios the redshift evolution we get is 
very small, which is consistent with observations.

\begin{figure}[htb]

\centerline{\epsfxsize=8 cm \epsfysize=5.5 cm \epsfbox{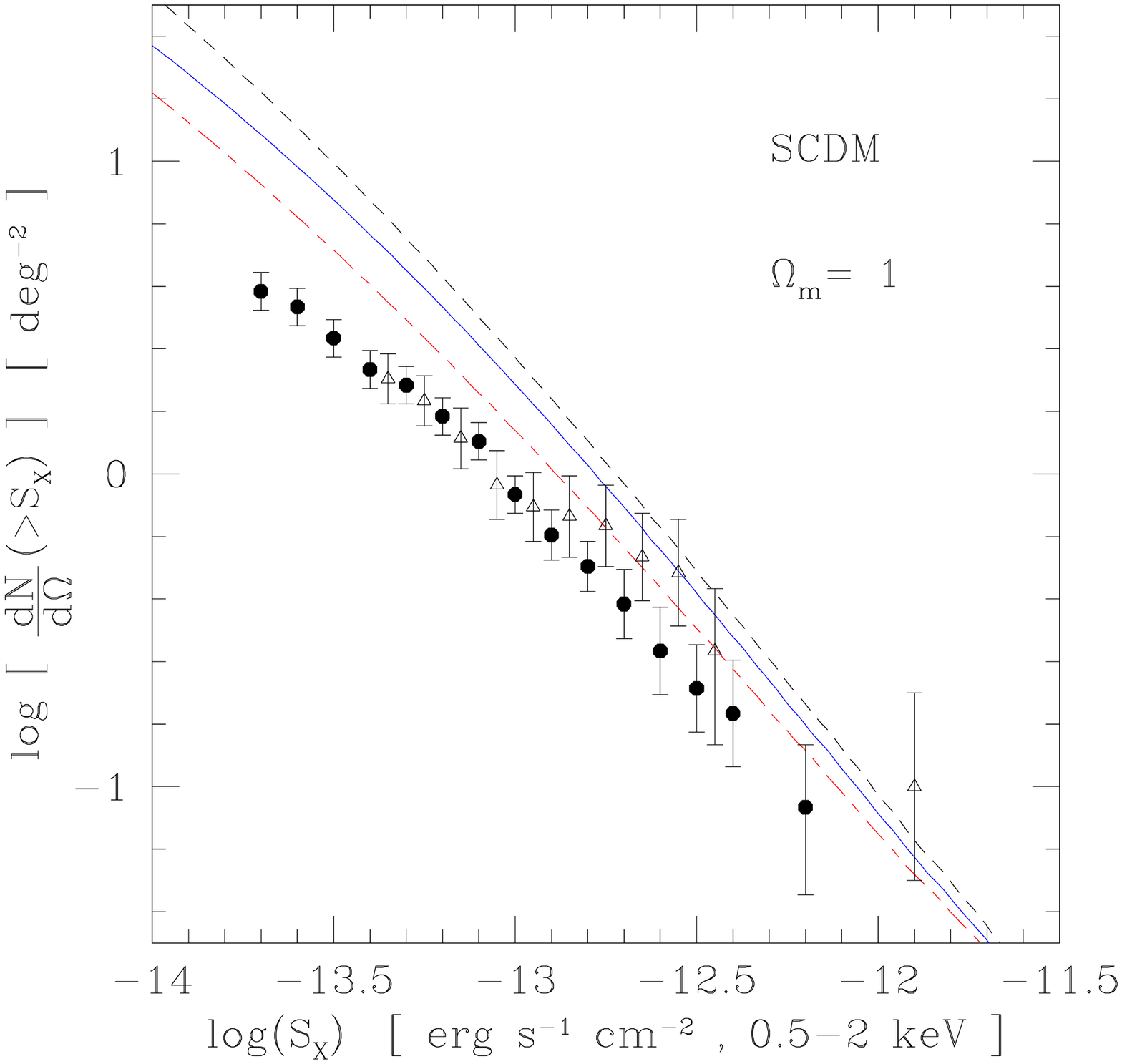}}
\centerline{\epsfxsize=8 cm \epsfysize=5.5 cm \epsfbox{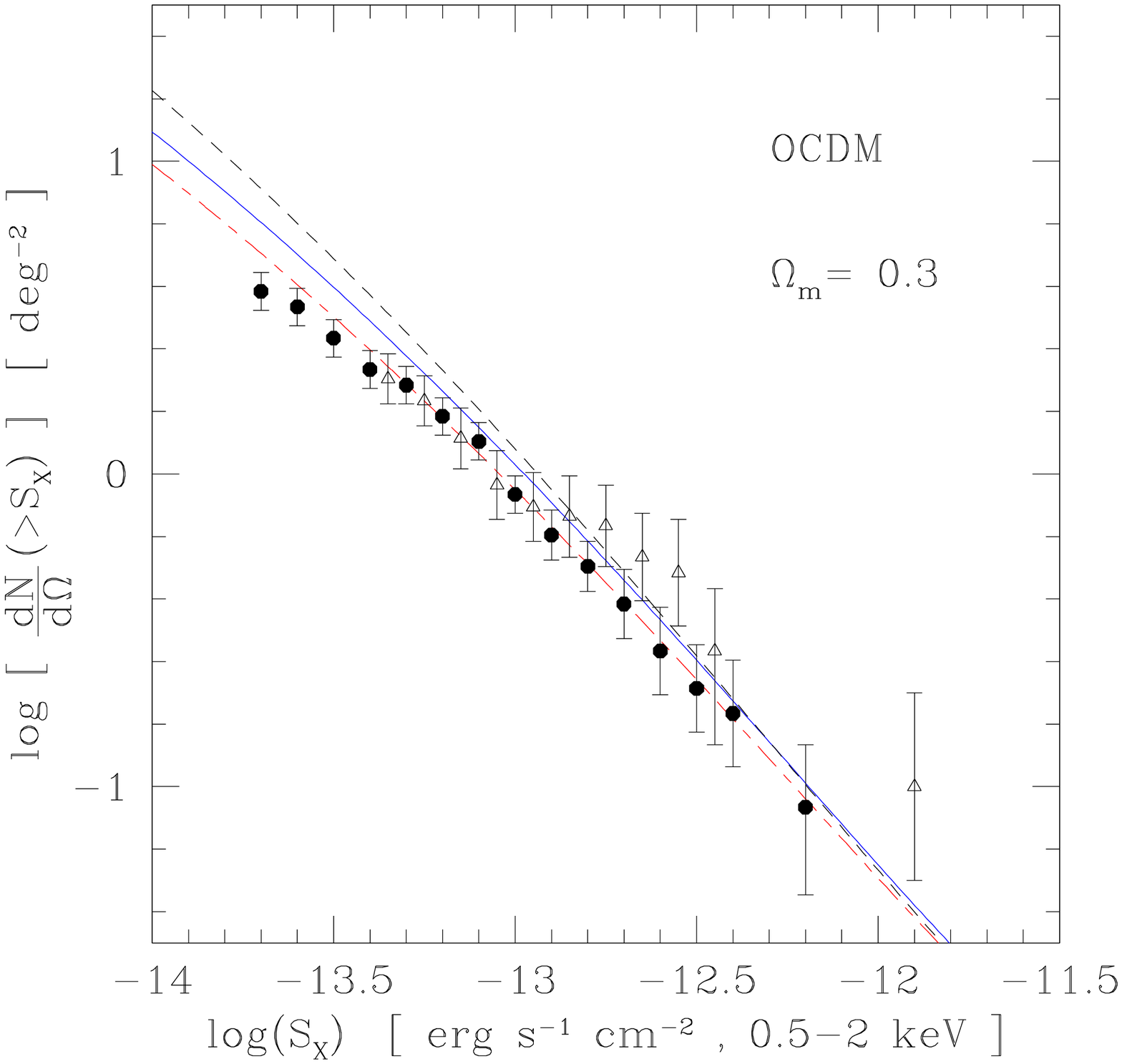}}

\caption{The number of clusters per square degree brighter than an X-ray flux 
limit $S_X$. The solid line corresponds to the scaling formulation and the 
dashed lines to the PS prescription. The low dot-dashed curve is a no-evolution 
model. The data points are from Jones et al. (1998) (triangles) and Rosati et 
al. (1998) (disks). The upper (resp. lower) panel corresponds to the SCDM (resp. 
OCDM) scenario.}
\label{figcountSXclus}

\end{figure}

We display in Fig.\ref{figcountSXclus} the integrated cluster surface density 
above an X-ray flux threshold $S_X$ in the frequency band $0.5-2$ keV:
\beq
\frac{d{\cal N}}{d\Omega} (>S_X) = \int dz \frac{dV}{d\Omega dz} 
\int_{M_i'}^{\infty} \eta(M,z) \frac{dM}{M} .
\label{SX}
\eeq
The cutoff $M_i'(S_X,z)$ corresponds to the X-ray flux $S_X$. It is obtained 
from the temperature-luminosity relation and the flux-luminosity relation:
\beq
S_X = \frac{L_X}{4 \pi r_{lum}(z)^2} ,
\label{SX}
\eeq
where the distance $r_{lum}(z)$ is the luminosity distance 
up to redshift $z$. Here 
the luminosity $L_X$ is obtained as in (\ref{LX}) with $\nu_i(z)=\nu_i (1+z)$ to 
take into account the redshift of the observed frequency band $0.5-2$ keV. Our 
results agree with observations, although our normalization is slightly
 too high for 
the critical density universe. In both panels, the dot-dashed curve corresponds 
to a no-evolution model where the comoving cluster luminosity function does not 
vary with $z$ and remains equal to its value at $z=0$, shown in 
Fig.\ref{figetaLclus}. Note that evolution effects are not very large. The 
scaling and PS approaches give very close results, although we can still 
recognize that the PS approximation predicts fewer massive objects but more 
small halos. Indeed, this difference is somewhat smeared out in 
Fig.\ref{figcountSXclus} because the X-ray sources seen with a given flux $S_X$ 
correspond to a large variety of objects located at different redshifts.

\section{Galaxies and quasars versus groups and clusters}
\label{Galaxies and quasars versus groups and clusters}

\subsection{Galaxy X-ray luminosity function}
\label{Galaxy X-ray luminosity function}

In addition to clusters, galaxies may also emit in X-rays when they form. 
Indeed, in order to make stars and build a galaxy the gas embedded within a dark 
matter halo must cool and fall into the gravitational potential well. During 
this process, the gas can radiate some energy in the X-ray band by 
bremsstrahlung, especially for the most massive galaxies with a large virial 
temperature $T \sim 10^7$ K $\sim 1$ keV. As a consequence, some of the X-ray 
sources one could observe on the sky may be high-redshift newly-born galaxies. 
Note that there will be obvious  observational difficulties
to distinguish X-ray emitting galaxies from small groups 
of similar mass,
containing  a few smaller galaxies.
 
To derive this galaxy X-ray luminosity function that is to be compared to the one for groups and rich clusters, we first 
need the galaxy multiplicity function.

\subsubsection{Galaxy multiplicity function}
\label{Galaxy multiplicity function}

\begin{figure}[htb]

\centerline{\epsfxsize=8 cm \epsfysize=5.5 cm \epsfbox{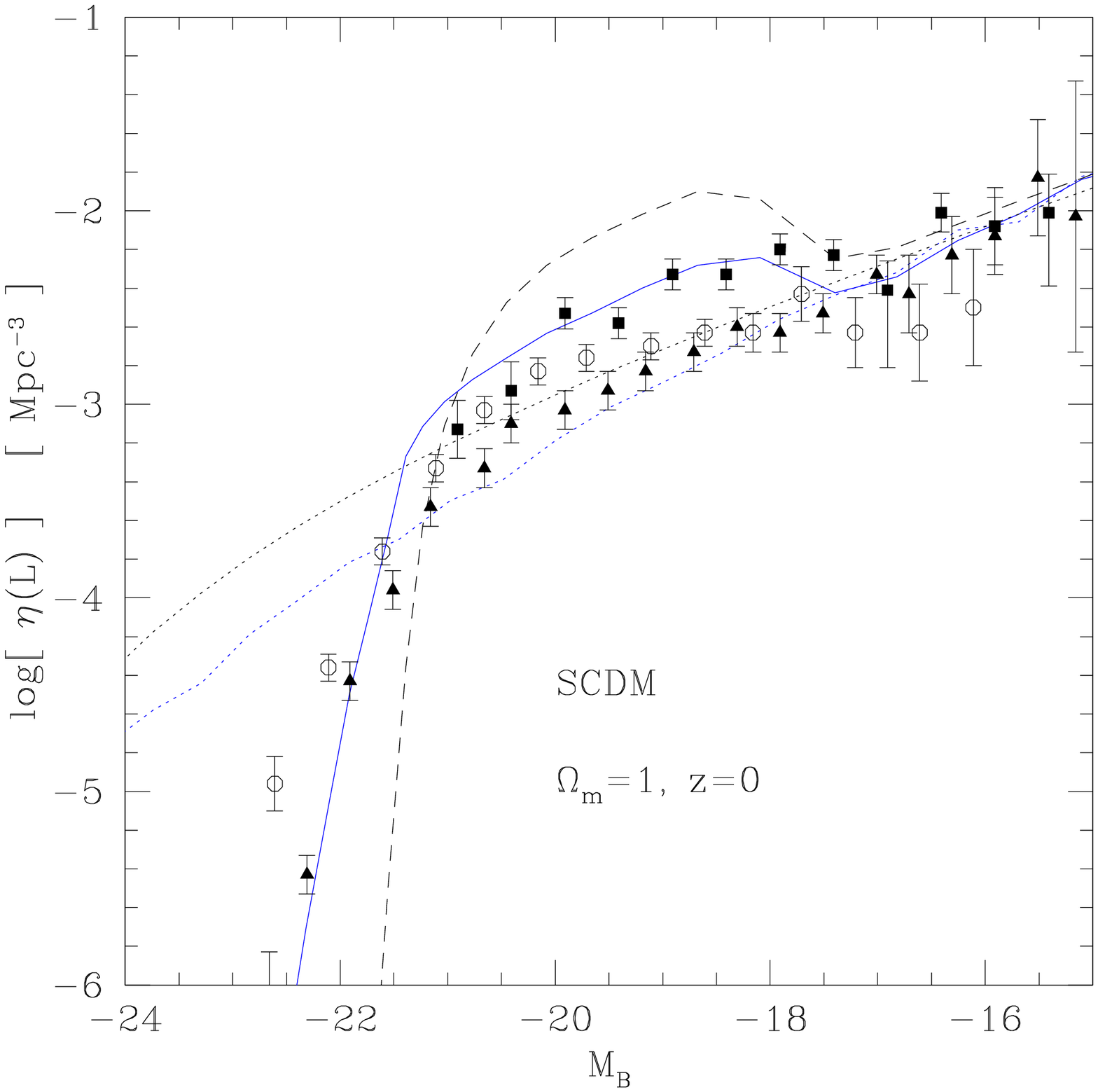}}
\centerline{\epsfxsize=8 cm \epsfysize=5.5 cm \epsfbox{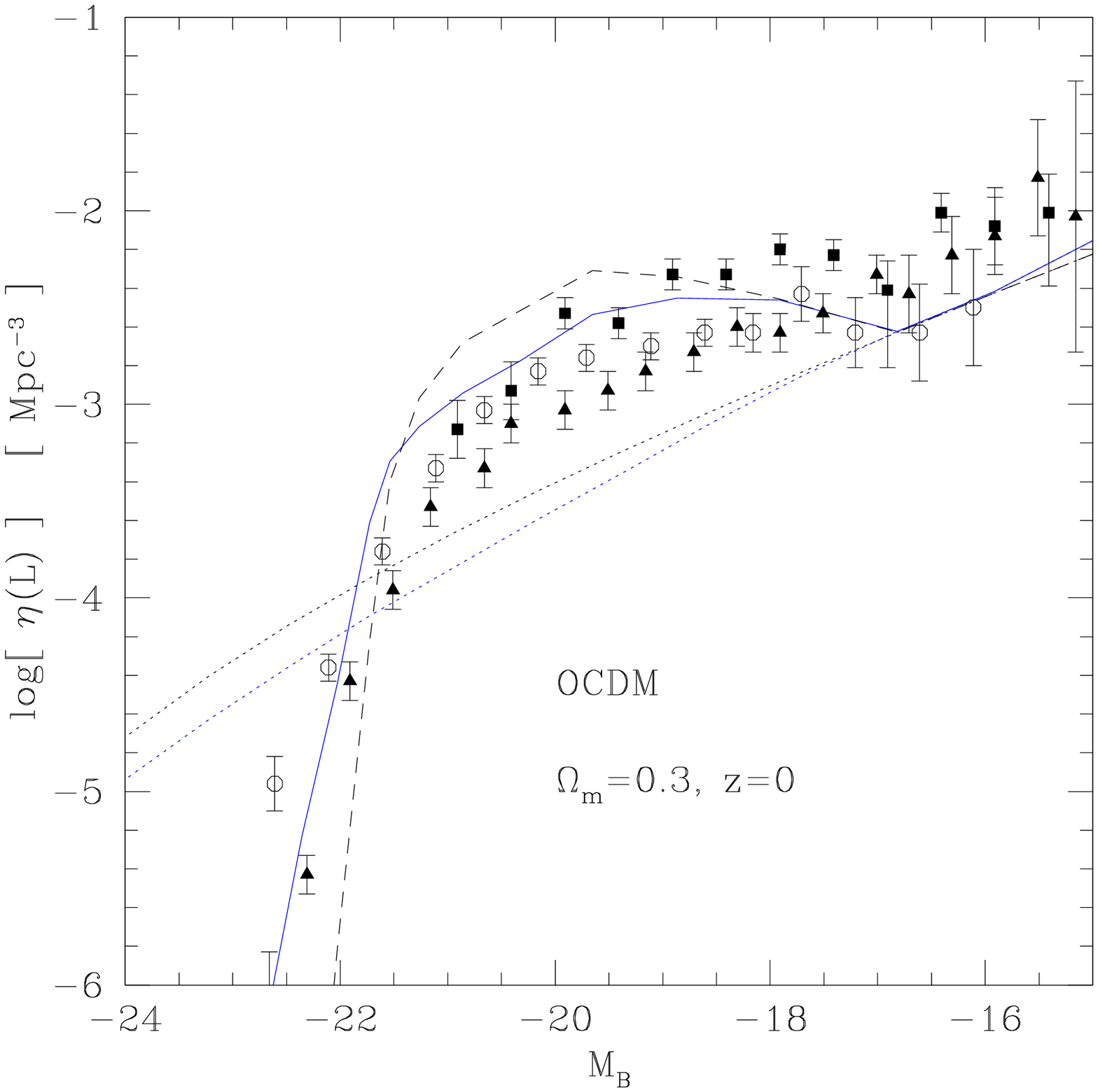}}

\caption{The comoving galaxy B-band luminosity function at $z=0$. The solid 
lines show our scaling approach and the dashed-lines a modified PS prescription. 
The dotted lines correspond to the scenario (C) where galaxies are defined by a 
constant density threshold $\Delta_c(z)$, for both prescriptions (thus the upper 
dotted curve is the standard PS mass function). The data points are observations 
from Loveday et al. (1992) (circles), Ellis et al.(1996) (filled squares) and 
Zucca et al. (1997) (triangles). The upper (resp. lower) panel corresponds to the 
SCDM (resp. OCDM) cosmology.}
\label{figgal}

\end{figure}

We use the galaxy formation model described in detail in Valageas \& Schaeffer 
(1999). This previous study is consistent with the present work (all parameters 
have the same values) and it was checked against observations for various galaxy 
properties. In particular, it is based on the same formalism described in 
Sect.\ref{Formalism} to derive the mass functions of dark matter halos. We 
briefly recall here the characteristics of this model we need for our purposes. 
We define galaxies by the requirement that two constraints be satisfied by the 
underlying dark matter halo: i) {\it a virialization condition} 
$\Delta>\Delta_c$ and ii) {\it a cooling constraint} $t_{cool} < t_{H}(z_{form})$ 
which states that the gas must have been able to cool within a few Hubble times 
at formation. We assume that the Hubble time at formation is given by the 
dynamical time, i.e. that the density of these dark matter halos does not evolve 
much after collapse. Note that just-collapsed halos ($\Delta=\Delta_c$) satisfy 
$t_{dyn} \sim t_H$ by definition. Thus, the dark matter radius of the halos we 
identify with galaxies is given by the conditions:
\beq
t_{cool} < t_{dyn} \hspace{1cm} \mbox{and} \hspace{1cm} \rho > (1+\Delta_c) 
\rhoa .
\label{sys1}
\eeq
In other words, the radius $R(T,z)$ of galaxies of virial temperature $T$ at 
redshift $z$ is:
\beq
R(T,z) = \mbox{Min} \left( R_{cool} , R_{vir} \right)
\eeq
where $R_{cool}$ is a ``cooling radius'' defined by $t_{cool} = t_{dyn}$ while
$R_{vir}$ is the ``virial radius'' defined by $\Delta=\Delta_c$. 
Note that
for clusters, 
 by 
definition, 
we have $R=R_{vir}$. At low redshift, for small masses, cooling is very efficient so that the virialization condition is 
the most constraining one in (\ref{sys1}). Hence $R=R_{vir}$ and these galactic 
halos are defined by the usual density contrast threshold $\Delta_c(z)$. On the 
other hand, for large masses (i.e. high $T$) cooling is inefficient and only occursfor high gas densities, so that galactic halos 
are defined by $R=R_{cool}$. This means that their mean density contrast is 
larger than $\Delta_c$ and that these objects formed at a larger redshift than 
the one 
we would have obtained by considering
just-collapsed objects defined by 
$\Delta=\Delta_c$. At large $T$, where bremsstrahlung is the main cooling 
process, the cooling radius goes over  to a constant $R_{cool} \sim 100$ kpc. 
Finally, we use a simple star formation model which takes into account infall 
into the inner parts of the galaxy, feedback from supernovae (proportional to 
$1/T$ for small galaxies as in Kauffmann et al. 1993) and with a star formation 
time-scale proportional to the dynamical time (see Valageas \& Schaeffer 1999 
for details). 

We recall in Fig.\ref{figgal} the B-band luminosity function we obtain at $z=0$. We can check that our predictions agree reasonably well with observations. 
Moreover, we can check that the ``extended PS'' prescription (dashed lines) 
predicts more intermediate mass halos and fewer very massive objects than the 
scaling mass function, as noted in Sect.\ref{Formalism}. This is consistent 
with the behaviour we already found
for clusters in Sect.\ref{Evolution of the mass 
function}. In addition, in order to clearly show the importance of correctly 
identifying the galactic halos we also display in Fig.\ref{figgal} the 
luminosity functions we would obtain (dotted curves) for a ``model'' (C) where 
all halos are defined by the virialization constraint (constant density 
contrast):
\beq
\mbox{(C)}: \; \Delta_{gal}(x) = \Delta_c(z) \hspace{0.6cm} \mbox{for all} 
\hspace{0.3cm} x
\label{C}
\eeq
with all other parameters (i.e. the star formation model) kept unchanged. Of 
course, this ``model'' is only shown for illustrative purposes since it is 
clearly inadequate for massive halos. Indeed, in the case (C) at large masses we 
identify clusters or groups and not galaxies! Thus, we can see in 
Fig.\ref{figgal} that for faint luminosities which correspond to small halos, the 
``model'' (C) superposes onto our actual galaxy model while for large 
luminosities it leads to huge galaxies which are not observed.
 In particular, it implies a bright magnitude tail of the 
luminosity function which is much too flat. Note that Monaco et al. (2000), 
using a model for quasars similar to ours (see below Sect.\ref{Quasar 
multiplicity function}), also manage to recover observations for both galaxy and 
quasar luminosity functions. However, in a standard procedure that
has been used  (Schaeffer \& Silk, 1988)
 as soon as the PS prescription became popular,
in order to correct the PS mass function 
so as to use it for galactic halos they multiply the PS prediction 
by a factor 
$\exp[-(M/M_{cool})^{4/3}]$
which is fitted to the cutoff of the observed galaxy 
luminosity function.
 Although this procedure can improve the mass 
functions it does not really deal with the ``subclustering problem'' itself 
(each massive halo still corresponds to a cluster). In contrast, our 
prescription has the serious advantage of taking into account this 
``subclustering problem'' in a very natural fashion using physical arguments 
based on cooling conditions, (see (\ref{sys1})), and to count the individual 
galactic halos themselves so that one can study their internal properties. 
Moreover, it is interesting to note that in addition to a strong cutoff of the 
luminosity function the use of the proper constraints (\ref{sys1}) leads to a 
non-trivial ``plateau'' in the range $-21 \la M_B \la -16$ which provides good 
agreement with observations.

\subsubsection{Galaxy X-ray emission}
\label{Galaxy X-ray emission}

\begin{figure}[htb]

\begin{picture}(230,320)
\epsfxsize=12 cm
\epsfysize=12 cm
\put(20,-10){\epsfbox{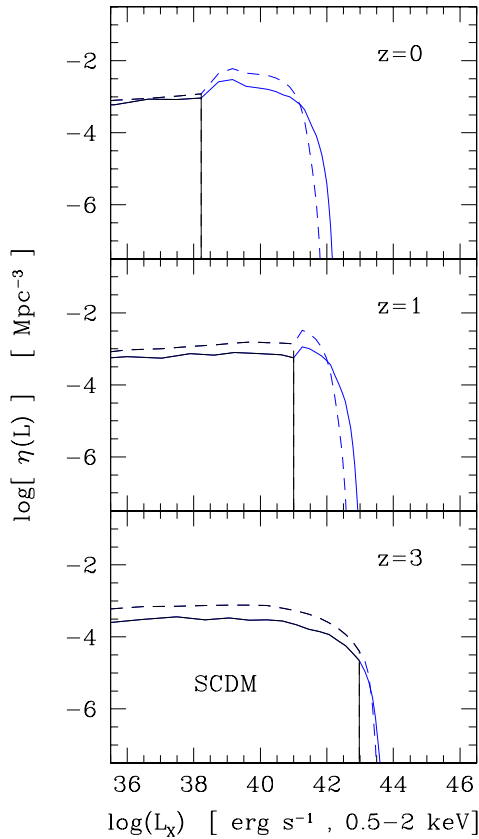}}
\end{picture}

\caption{The comoving galaxy X-ray luminosity function in the rest-frame 
frequency band $0.5-2$ keV for the SCDM case. The solid line corresponds to the 
scaling formulation and the dashed line to the PS prescription, for the ``Hot'' 
scenario. The vertical line shows the cutoff at $T_f(z)$. For the ``Cold'' model 
the luminosity function vanishes at higher luminosities and it is equal to the 
``Hot'' model prediction at fainter luminosities.}
\label{figLXO1}

\end{figure}

\begin{figure}[htb]

\begin{picture}(230,320)

\epsfxsize=12 cm
\epsfysize=12 cm
\put(20,-10){\epsfbox{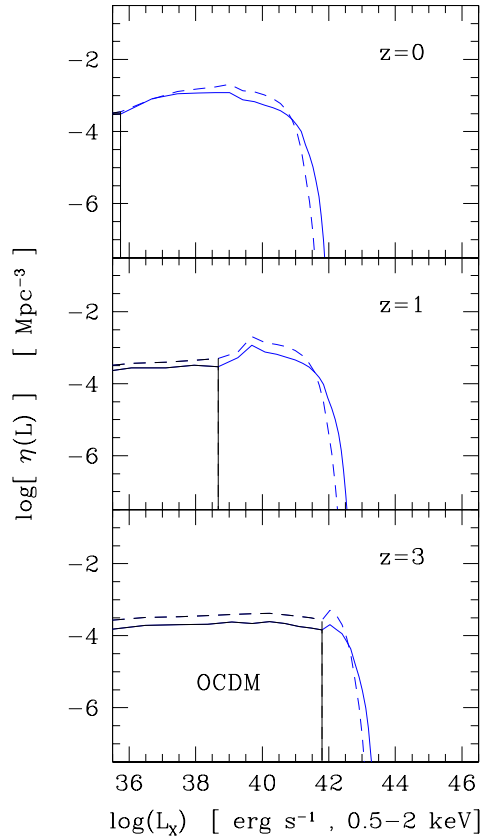}}

\end{picture}

\caption{The comoving galaxy X-ray luminosity function in the rest-frame 
frequency band $0.5-2$ keV for the OCDM scenario. Same curves as in 
Fig.\ref{figLXO1}.}
\label{figLXO03}

\end{figure}

As explained above galaxies are defined by the two constraints (\ref{sys1}). 
Hence at any redshift there is a characteristic virial temperature $T_f(z)$ 
which marks the transition between the low temperature regime, where 
$R=R_{vir}$, and the high temperature regime, where $R=R_{cool}$. It increases 
with $z$ and at $z=0$ we have $T_f \sim 10^6$ K. In this approach, large objects 
defined by $\Delta=\Delta_c$ with a virial temperature $T>T_f(z)$ are made of 
several galaxies and correspond to groups or clusters. They can be subdivided 
into several subunits which verify the constraints (\ref{sys1}). Of course, 
halos with a virial temperature which is only slightly higher than $T_f(z)$, merely consist of one galaxy with some gas falling from its surroundings which 
have not cooled yet. However, we shall identity just-collapsed halos above 
$T_f(z)$ as groups or clusters, which should provide a reasonable estimate of 
the transition, while we shall call just-collapsed halos below $T_f(z)$, galaxies. In particular, the mass functions of groups and clusters we used above 
were set to 0 for $T<T_f(z)$ (this also corresponds to the cutoff $M_{i}(z)$ 
which appeared in (\ref{y3}) to obtain $\lag y \rag$). However, this has no 
effect on the quantities we have studied so far because we considered high virial 
temperatures $T>0.5$ keV at low redshifts $z<1$. Since in our model galaxies 
below and above $T_f$ have a rather different history we consider each regime 
separately.

First, we write the galaxy luminosity (due to the  cooling of the gas) as:
\beq
L_{bol} = \frac{E}{t_{cool}} = \frac{3}{2} \frac{M_{hot} k T}{\mu m_p t_{cool}}
\label{Lbolgal}
\eeq
where $M_{hot}$ is the mass of hot gas (i.e. of the order of the virial 
temperature $T$) and $t_{cool}$ is its cooling time. Galaxies with $T<T_f(z)$ 
are defined as ``just-collapsed'' objects ($\Delta=\Delta_c$) and they satisfy 
$t_{cool} < t_H(z)$. Although the latter, by definition,  
have undergone a  major merging event in their recent past
(at a redshift $z+\delta z$ with $\delta z/(1+z) \ll 1$), all of them are not necessarily in the 
midst of such a process. The time which has elapsed since the last merging, is 
measured by $t_H $, so we write for the galaxy X-ray luminosity function
\beq
T < T_f : \hspace{0.3cm} \eta(L_{bol}) \frac{dL_{bol}}{L_{bol}} = 
\frac{t_{cool}}{t_H} \; \eta(M) \frac{dM}{M} .
\label{galetaLbol1}
\eeq
This ensures that we only count the galaxies where the gas has not had time to 
cool significantly since the last major merging event (while we neglect the 
contribution of the small objects which had time to cool). Then, the mass 
$M_{hot}$ of hot gas characteristic of these objects is of the order of the total 
mass of gas $M_g$, hence we take $M_{hot}=M_g$ (and the cooling time is evaluated 
using the gas density $\rho_g = (1+\Delta_c) \rhoa_b$). On the other hand, 
massive old galaxies with $T>T_f$ have already had time to cool since they 
satisfy $t_{cool} = t_H(z_{form}) < t_H(z)$. However, if cooling is 
inhomogeneous and there is an approximate pressure equilibrium, a diffuse gaseous hot component may be present in the galactic halo with a density given by the 
condition $t_{cool}=t_H(z)$, as we discussed in Sect.\ref{Cooling radius} for 
clusters. This component could be a left-over of the initial baryonic content of 
the galaxy (which gradually cools with time and falls into the inner parts of 
the galaxy to form stars) but it could also be replenished by the ejection of 
matter from the center of the galaxy by supernovae. In order to investigate the 
range of the galactic contribution to the overall X-ray emission we consider the 
following two models. First, we assume that there is no hot component (or it is 
negligible) so that in this ``Cold'' scenario old galaxies with $T>T_f(z)$ no longer emit in X-rays:
\beq
\mbox{(Cold)} : \hspace{0.2cm} L_{bol}=0 \hspace{0.4cm} \mbox{for} 
\hspace{0.4cm}  T > T_f(z) .
\label{cold}
\eeq
This implies a sharp cutoff for the galaxy X-ray luminosity function at the 
luminosity which corresponds to the transition $T_f$ (i.e. there is an upper bound 
for $L_{bol}$). Note however that these galaxies contribute to the X-ray 
luminosity function at the redshift $z_{form}$, when they formed. Second, we 
consider a ``Hot'' model where there is a diffuse hot component at the virial 
temperature $T$ with a characteristic density:
\beq
\mbox{(Hot)} : \hspace{0.2cm} \rho_{hot} = \frac{3 \mu_1^2 m_p k T}{2\mu 
\Lambda_c(T) t_H} \hspace{0.4cm} \mbox{for} \hspace{0.4cm} T > T_f(z) .
\label{hot}
\eeq
This gives the mass $M_{hot}$ at  the galactic radius $R$ and the luminosity 
$L_{bol}$ from (\ref{Lbolgal}) where we now have $t_{cool}=t_H(z)$. Then, we 
obtain the X-ray luminosity function from the galaxy multiplicity function 
described in Sect.\ref{Galaxy multiplicity function} as: $\eta(L_{bol}) 
dL_{bol}/L_{bol} = \eta(M) dM/M$. In practice, we can expect the actual X-ray 
luminosity function in our universe to be somewhere in between these two ideal 
scenarios. In particular, since the cooling radius is equal to the actual radius 
$R$ of the galaxy (by definition), there is not necessarily a large reservoir of 
hot gas at radii larger than $R_{cool}$ to ensure that the hot phase occupies all the volume 
 of the galactic potential well, in contrast to the case encountered for 
clusters. However, there may be some infall from the surrounding IGM (in 
addition to SNe ejecta). Hence the ``Hot'' scenario can be interpreted as an 
upper bound for the X-ray galaxy counts and the ``Cold'' scenario as a lower 
bound. Finally, we obtain the X-ray luminosity $L_X$ in a given frequency band 
$\nu_1 - \nu_2$ as in (\ref{LX}), which allows us to derive the galaxy 
luminosity function in this band.

We show in Fig.\ref{figLXO1} and Fig.\ref{figLXO03} the galaxy X-ray luminosity 
functions we obtain in this way for both cosmologies in the rest-frame frequency 
band $0.5-2$ keV. The solid lines (resp. dashed lines) correspond to the scaling 
formulation (resp. the PS prescription) for the ``Hot'' scenario (\ref{hot}). 
For the ``Cold'' model, the luminosity function vanishes at high luminosities 
above the cutoff $T_f(z)$ shown by the vertical solid line while it is equal to 
the ``Hot'' model prediction at fainter luminosities. As explained above the 
cutoff $T_f(z)$ increases with $z$ which leads to a larger luminosity cutoff 
$L_f(z)$ at higher $z$. Below $T_f$ the prefactor $t_{cool}/t_H$ diminishes the 
contribution of small and faint galaxies since an increasingly large fraction of 
these objects is cold. Moreover, the contribution of small galaxies (which have 
a low virial temperature $T$) is strongly suppressed by the factors $\exp(-h_P 
\nu/k T)$ which enter the X-ray luminosity (\ref{LX}). 

As usual, the redshift evolution is smaller for the open universe (see Valageas 
\& Schaeffer 1999) as the galaxy multiplicity function evolves more slowly (for 
the same reason as for the cluster multiplicity function). However, the cutoff 
$L_{Xf}(z)$ increases faster with $z$ than for the SCDM case. This is due to the 
fact that the temperature attached to these galaxies (in particular $T_f$) is 
somewhat smaller for the low-density universe which implies that the factors 
$\exp(-h_P \nu/k T)$ are more sensitive to $T_f$ and evolve faster with $z$. 
Note that the X-ray luminosity $L_{Xf}(z)$ is indeed lower for the open 
universe. As explained in Sect.\ref{Galaxy multiplicity function} the cooling 
constraint plays a greater role at small redshift, as shown by the comparison 
between the ``Hot'' and ``Cold'' models. This implies that the difference 
between both scenarios is largest at $z=0$. Note that even for the ``Hot'' model 
the high luminosity cutoff is stronger at lower $z$. This is due to the gradual 
decline with time of the mass of hot gas $M_{hot}$ attached to these halos 
together with the larger Hubble time $t_H(z)$, see (\ref{hot}) and 
(\ref{Lbolgal}).

\subsection{Quasar X-ray luminosity function}
\label{Quasar X-ray luminosity function}

In addition to galaxies, groups and clusters, quasars are 
X-ray emitters.
 In fact, as we shall see in the next section their contribution 
to the X-ray flux is much larger than the galactic emission because of their 
harder radiation spectrum (which is roughly similar to a power-law as opposed to 
the black-body spectrum of stars). Thus, it is of interest to estimate the 
quasar source counts since they dominate at fluxes smaller than those 
corresponding to clusters. Moreover, it allows us to obtain a complete 
description of X-ray objects within the framework of a unified model. Indeed, 
since in our model QSOs correspond to galaxies where some gas is accreted by a 
central black hole the quasar luminosity function is derived from the galaxy 
multiplicity function. Hence it is fully consistent with the cluster and galaxy 
mass functions we obtained in the previous sections. We now briefly describe our 
model for quasars, which is similar to the one used in Valageas \& Silk 
(1999a,b) to study the reheating and reionization history of the universe.

\subsubsection{Quasar multiplicity function}
\label{Quasar multiplicity function}

First, we assume that the mass $M_Q$ which is available to fuel the quasar is 
proportional to the sum of the mass of central gas $M_{gc}$ and of stars which 
formed lately $\Delta M_s$:
\beq
M_Q = F ( M_{gc} + \Delta M_s ) .
\label{MQ}
\eeq
Thus, at late times when most of the gas is consumed into stars the total mass 
of the black hole is $M_{BH} \sim F M_s$. We use $F=0.005$ since observations 
from Magorrian et al. (1998) find that $M_{BH} \sim 0.005 M_{sph}$ where 
$M_{sph}$ is the mass of the stellar bulge. From the model of star formation 
described in Valageas \& Schaeffer (1999) we have: 
\beq
M_{gc} = \left( 1+\frac{T_{SN}}{T} \right)^{-1} e^{-\lambda} M_b
\label{Mgc}
\eeq
and
\beq
\Delta M_s \equiv M_s \; \min \left( 1 , \frac{t_H}{M_s} \frac{dM_s}{dt} \right) = 
\lambda \; e^{-\lambda} M_b ,
\label{Ms}
\eeq
where $M_b$ is the mass of baryons in the galaxy and 
as a function of our scaling parameter $x$
\beq
\lambda(x) = \frac{p}{\beta_d} \left(1+\frac{T_{SN}}{T} \right)^{-1} \sqrt{ 
\frac{(1+\Delta)_{gal}(x)}{(1+\Delta_c)} } .
\label{lambdax}
\eeq
Here $p/\beta_d \simeq  0.5$ is a parameter which enters the definition 
of the dynamical time, while $T_{SN} \sim 10^6$ K describes the ejection of gas 
by supernovae and stellar winds (see also Kauffmann et al. 1993). The coefficient 
$\lambda$ measures the efficiency of star formation. Thus, small galaxies with 
$T \ll T_{SN}$ have $\lambda \ll 1$ because supernovae eject a large fraction of 
the gas out of the inner parts of the galaxy. Hence $M_{gc} \ll M_b$ and $\Delta 
M_s \sim M_s \ll M_b$ as seen in (\ref{Mgc}) and (\ref{Ms}). On the other hand, 
massive old galaxies with a density contrast $\Delta_{gal} > \Delta_c$ have a 
small dynamical time, hence a small star formation time in our model. This leads 
to the factor $\sqrt{(1+\Delta)_{gal}}$ in (\ref{lambdax}). Hence they have 
$\lambda \gg 1$ and they have already consumed most of their gas, so that the 
quasar runs out of fuel, see (\ref{Mgc}) and (\ref{Ms}). Next, if we note $\fEd$ 
the Eddington ratio we write the quasar luminosity $L_Q= \fEd \LEd$ as:
\beq
L_Q = \fEd \frac{M_Q \; c^2}{t_*} \hspace{0.5cm} \mbox{with} \hspace{0.5cm} 
t_*=4.4 \; 10^8 \mbox{yr} ,
\label{LQ}
\eeq 
while the quasar life-time $t_Q$ is:
\beq
t_Q = \frac{\epsilon_Q}{\fEd} \; t_* ,
\label{tQ}
\eeq
where $\epsilon_Q=0.05$ is the quasar radiative efficiency. Finally, we write 
the quasar multiplicity function we would obtain without any scatter as:
\beq
\eta_Q(M_Q) \frac{dM_Q}{M_Q} = \lambda_Q \; \frac{t_Q}{t_M} \; \eta_g(M) 
\frac{dM}{M} 
\label{etaQ}
\eeq
where $\eta_g(M) dM/M$ is the galaxy mass function obtained in Sect.\ref{Galaxy 
X-ray luminosity function}. Here $t_M \sim t_H$ is the evolution time-scale of 
galactic halos of mass $M$ defined by:
\beq
t_M^{-1} = \mbox{Max} \left( t_H^{-1} \; , \; \frac{1}{\eta_g(M)} \; 
\frac{\pl}{\pl t} \eta_g(M) \right) .
\eeq
We use $\lambda_Q \sim 0.05$. Note that here we only have two free parameters 
$(\fEd F)$ and $(\lambda_Q \epsilon_Q/\fEd)$ so that we could
for instance  use a larger 
$\fEd$ with a smaller $F$ and larger $\lambda_Q$. Moreover, the 
properties of quasars  show a significant scatter. For instance, 
Magorrian et al. (1998) find that the decimal logarithm of the ratio 
$F=M_{BH}/M_{sph}$ obeys a Gaussian distribution of mean $-2.28$ and standard 
deviation $\sim 0.51$. Hence we assume here that the actual luminosity 
$L_{Qscat}$ of QSOs is related to the luminosity $L_Q$ defined in (\ref{LQ}) by:
\beq
L_{Qscat} = e^u \; L_Q \hspace{0.3cm} , \hspace{0.3cm} P(u) = 
\frac{1}{\sqrt{2\pi \sigma_u^2}} \; e^{-u^2/2\sigma_u^2}
\label{Pu}
\eeq
where $P(u)du$ is the probability distribution of the random variable $u$. Thus, 
we assume $P(u)$ to be a Gaussian of width $\sigma_u=1.15$ (resp. $\sigma_u=1$) 
for the SCDM (resp. OCDM) scenario. Note that $\sigma_u=1.15$ corresponds to the 
scatter observed by Magorrian et al. (1998) for the ratio $F$. Although part of 
this scatter may be due to the observational noise, which would decrease 
$\sigma_u$, the random variable $u$ also describes the scatter of the other 
properties of quasars like the Eddington ratio $\fEd$. Then, we obtain the final 
quasar luminosity function from (\ref{etaQ}) and (\ref{Pu}):
\beq
\eta_{Qscat}(L_{Qscat}) = \int_{-\infty}^{\infty} du \; P(u) \; \eta_Q \left( 
L_{Qscat} \; e^{-u} \right) .
\label{etaQscat}
\eeq

\begin{figure}[htb]

\begin{picture}(230,410)
\epsfxsize=26 cm
\epsfysize=16 cm
\put(-28,-30){\epsfbox{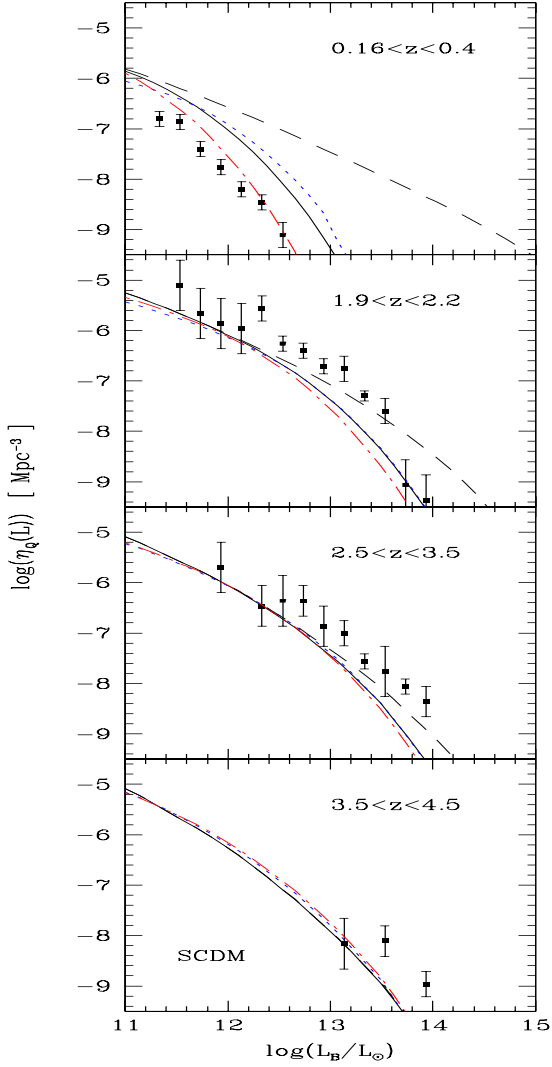}}
\end{picture}

\caption{The comoving quasar B-band luminosity function for the SCDM case. The 
data points are observations from Pei (1995). The solid lines correspond to our 
fiducial model, the dotted lines to (A), the dot-dashed lines to (B) and the 
dashed lines to (C), see main text.}
\label{figQLO1}

\end{figure}

\begin{figure}[htb]

\begin{picture}(230,410)
\epsfxsize=26 cm
\epsfysize=16 cm
\put(-28,-30){\epsfbox{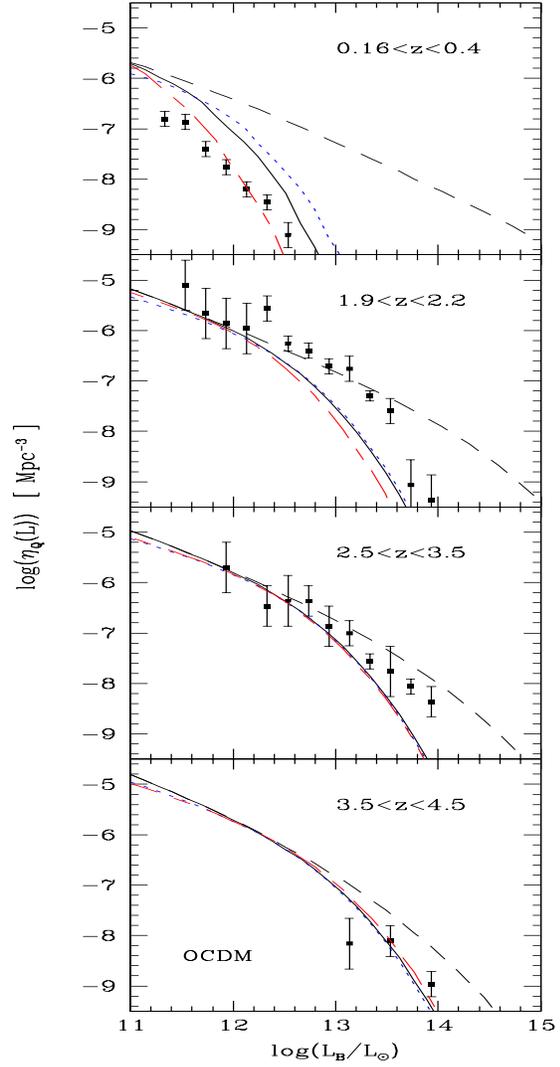}}
\end{picture}

\caption{The comoving quasar B-band luminosity function for the OCDM case as in 
Fig.\ref{figQLO1}.}
\label{figQLO03}

\end{figure}

Since observations suggest that the Eddington ratio $\fEd$ increases with the 
luminosity $L_Q$ (Padovani 1989; Wandel 1998) we use the parameterization:
\beq
\fEd = \mbox{Min} \left[ 2 \; , \; \left( \frac{\LEd}{10^{47} \; \mbox{erg/s}} 
\right)^{0.33} \right]
\label{fEd}
\eeq
in order to be consistent with the data obtained by Padovani (1989). This author 
finds that most of the observed dependence of $\fEd$ on redshift is accounted 
for by the dependence on luminosity (low $z$ quasars are fainter than high $z$ 
QSOs) hence we do not add any explicit dependence on redshift in our model (\ref{fEd}). However, in order to estimate the influence of these various 
parameterizations we also consider for the B-band luminosity functions displayed 
in Fig.\ref{figQLO1} and Fig.\ref{figQLO03} the following two alternative models 
(A) and (B) with:
\beq
\mbox{(A)}: \; \fEd = 1
\label{fEdA}
\eeq
and
\beq
\mbox{(B)}: \; t_Q = \mbox{Max} \left( 2 \times 10^8 \; (1+z)^{-1.5} \; 
\mbox{yr} \; , \; \frac{\epsilon_Q}{2} \; t_* \right) .
\label{fEdB}
\eeq
Thus, model (A) corresponds to a constant Eddington ratio of unity while model 
(B) corresponds to an accretion time which is roughly proportional to the Hubble 
time (with a lower bound so that $\fEd \leq 2$). Finally, in order to point out 
the importance of correctly identifying the galactic halos we also display in 
Fig.\ref{figQLO1} and Fig.\ref{figQLO03} the ``model'' (C) where all halos are 
defined by the virialization constraint as in (\ref{C}). Of course, as in 
Sect.\ref{Galaxy multiplicity function} we only show this ``model'' for 
illustrative purposes since it is not valid for massive halos. However, note 
that usual analytic models based on the PS approach (e.g., Haiman \& Menou 2000) 
define halos as in (C).

We show the B-band quasar luminosity functions that
we obtain in Fig.\ref{figQLO1} 
and Fig.\ref{figQLO03} for both SCDM and OCDM cosmologies. First, we note that 
our result (solid lines) agree reasonably well with observations over the whole 
range $0 < z < 4.5$. Moreover, we can check that both  models (A) 
(dotted lines) and (B) (dot-dashed lines) are quite close to our fiducial model.
 Thus, although our model is very crude our 
predictions should provide a reasonable estimate of the quasar multiplicity 
since they are not too sensitive to details. Note, however, that 
the model (B) shows a slightly better agreement with observations as it leads to a 
stronger decrease of the quasar number density at low $z$. This is due to the 
redshift evolution of the quasar life-time $t_Q$ which implies a smaller 
Eddington ratio $\fEd$ at low $z$. On the other hand, we can clearly see that 
the ``model'' (C) fails at low redshift $z < 2$. Indeed, since it counts 
groups or clusters as one halo it predicts a significant number of huge quasars, 
which is inconsistent with observations. Of course, at high redshift where the 
cooling condition (first constraint in (\ref{sys1})) plays no role the ``model'' 
(C) superposes onto our model (\ref{fEd}) since all galaxies are defined by the 
usual density threshold $\Delta_c(z)$.

\begin{figure}[htb]

\centerline{\epsfxsize=8 cm \epsfysize=8 cm \epsfbox{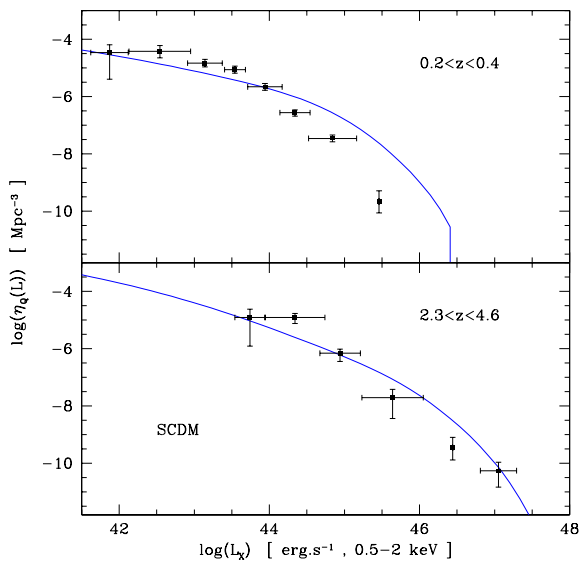}}

\caption{The comoving quasar X-ray luminosity function in the frequency band 
$0.5-2$ keV for the case $\Om=1$. The data points are observations from Miyaji 
et al. (1998).}
\label{figQLXO1}

\end{figure}

\begin{figure}[htb]

\centerline{\epsfxsize=8 cm \epsfysize=8 cm \epsfbox{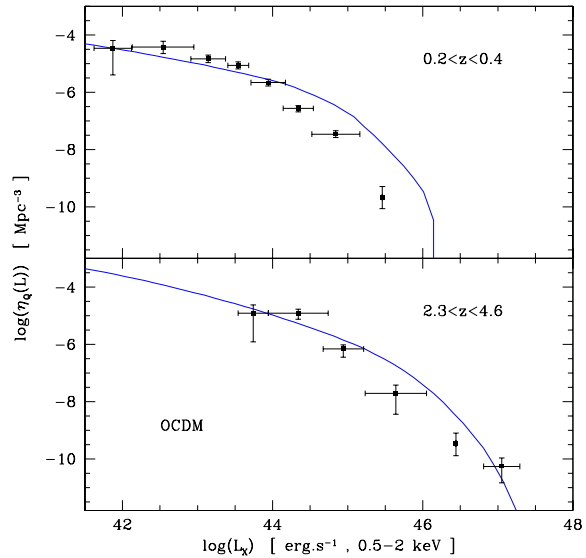}}

\caption{The comoving quasar X-ray luminosity function in the frequency band 
$0.5-2$ keV for the OCDM scenario.}
\label{figQLXO03}

\end{figure}

Thus, one cannot use the (standard) PS prescription to 
obtain the QSO luminosity function at low $z$. Hence one should be careful 
about the use of the PS mass function to derive the redshift evolution of quasar 
properties from the observed luminosity function. For instance, while Haiman \& 
Menou (2000) find that this procedure implies that the Eddington ratio $\fEd$ or 
the mass ratio $F$ must drop by a factor $\sim 100$ at low $z$ between $0 \la z 
\la 3$ (which would steepen the dashed curves shown in the upper panel in 
Fig.\ref{figQLO1} and Fig.\ref{figQLO03}), they note that a possible ``solution'' 
would be to correct the halo mass function in order to make sure that one counts galactic halos and not clusters. Indeed, our results show that the correct 
procedure (i.e. identifying galactic halos) provides by itself a reasonable 
agreement with observations, without any additional ad-hoc redshift dependence.
It could be tempting to introduce
 a redshift dependence of the form (\ref{fEdB}), to
 improve somewhat the fit to observations. Nevertheless,
we consider the present models 
too crude to allow one to draw such conclusions from the
observations: such a mouve would plug theoretical inaccuracies into an artificial ``observed'' evolution of (\ref{fEdB}) with redshift. Note however that 
such an evolution with redshift
may exist.

\subsubsection{Quasar X-ray emission}
\label{Quasar X-ray emission}

From the quasar multiplicity function we obtain the X-ray source counts which 
are identified as QSOs as in (\ref{SX}). We assume that the quasar spectrum is 
locally a power-law $L_{\nu} \propto \nu^{-1.5}$ around $\nu_1 = 1$ keV, 
normalized by $(L_1/L_{bol})=0.028$ with $L_1= \nu_1 L_{\nu}(\nu_1)$. We present 
in Fig.\ref{figQLXO1} and Fig.\ref{figQLXO03} the comoving quasar luminosity 
function we obtain for both SCDM and OCDM cosmologies. The frequency band 
$0.5-2$ keV corresponds to the observed spectrum (i.e. light was emitted between 
$0.5 (1+z)$ and $2 (1+z)$ keV). We see that we obtain a reasonable agreement 
with observations from Miyaji et al. (1998), both at low redshift $z \sim 0.3$ 
and high redshift $z \sim 3.5$. This could be expected from the results of 
Fig.\ref{figQLO1} and Fig.\ref{figQLO03} for the B-band luminosity function.

\subsection{Galaxy and quasar versus group and cluster counts}
\label{Galaxy versus group counts}

\begin{figure}[htb]

\centerline{\epsfxsize=8 cm \epsfysize=5.5 cm \epsfbox{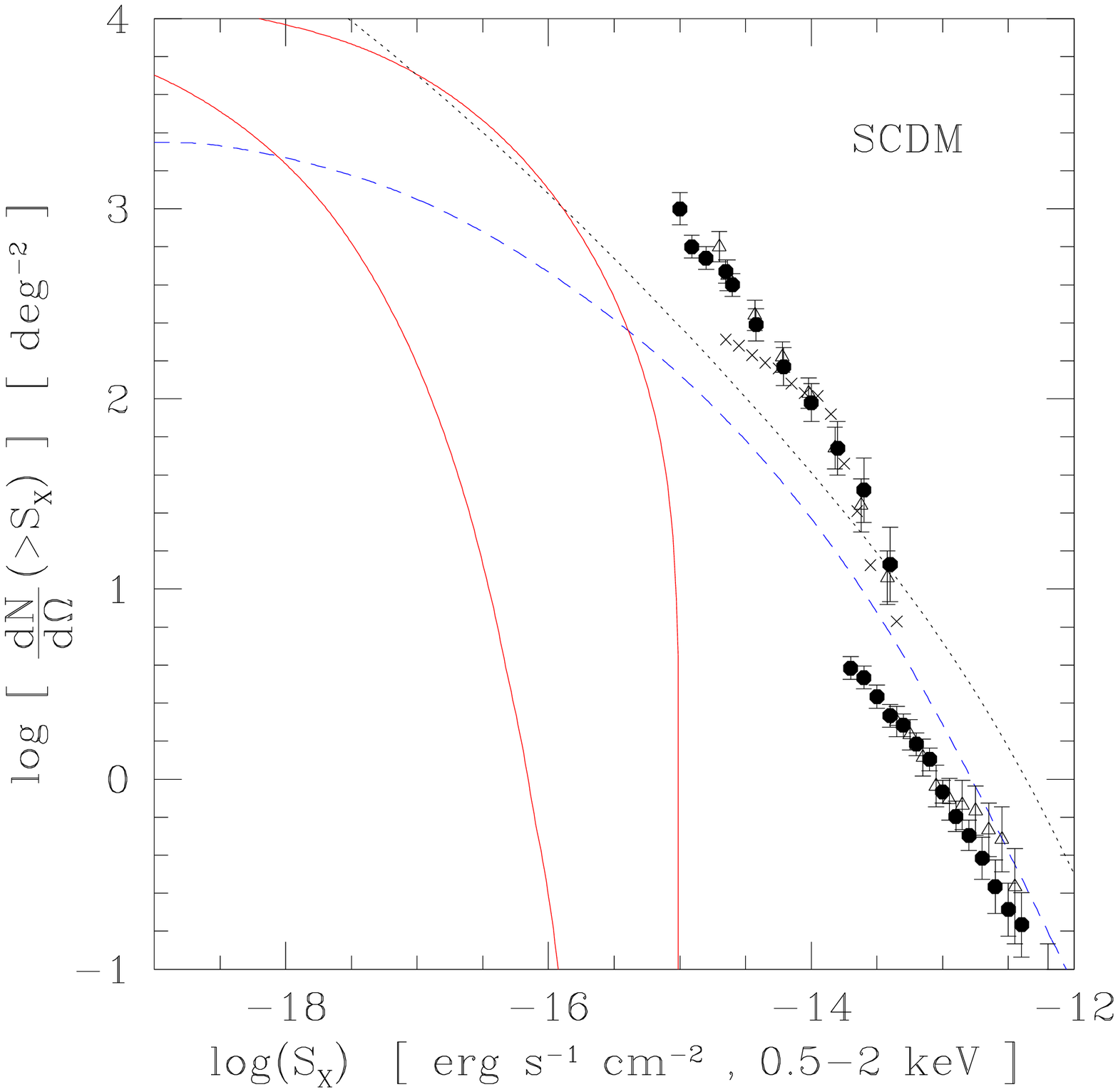}}
\centerline{\epsfxsize=8 cm \epsfysize=5.5 cm \epsfbox{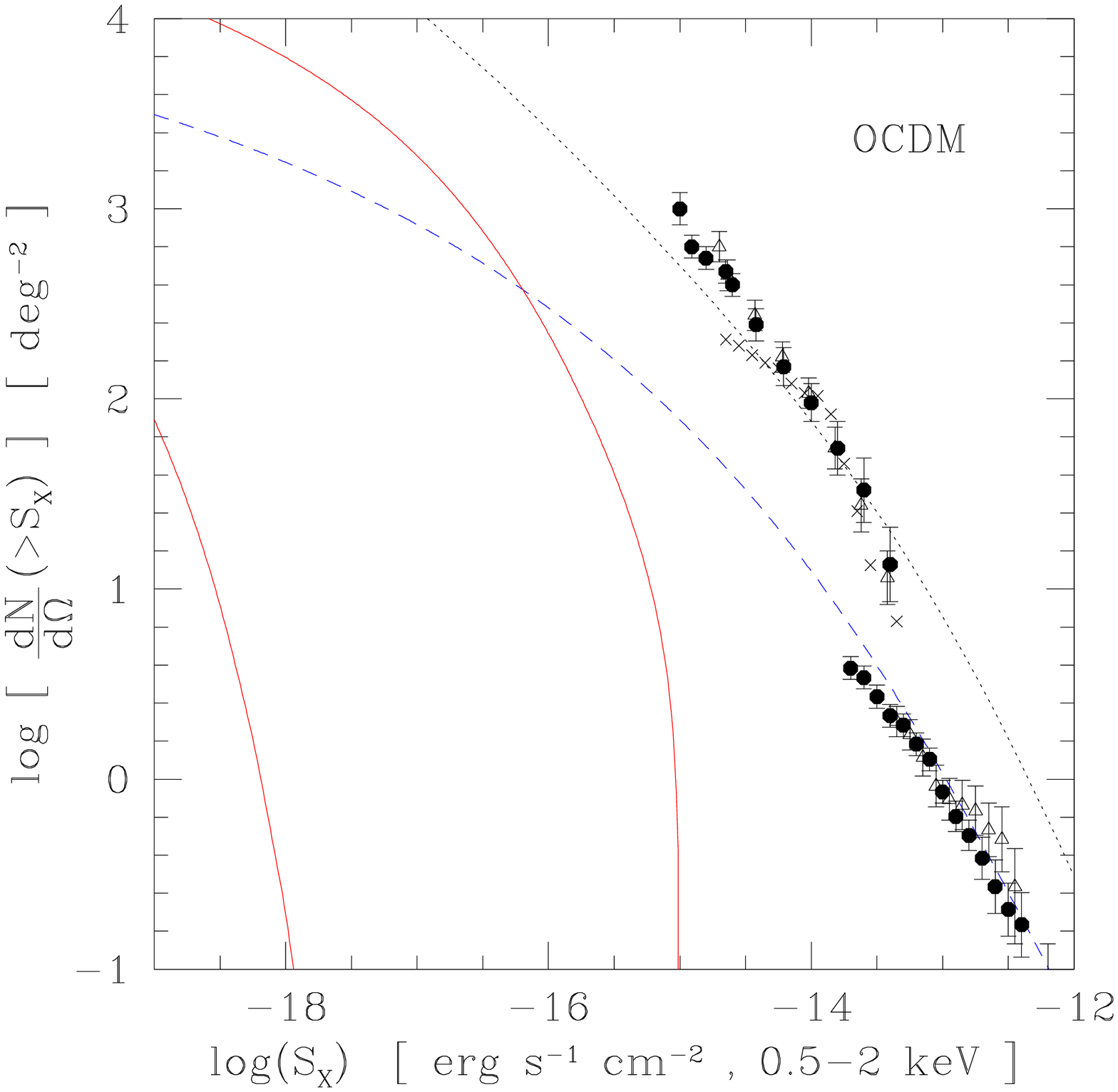}}

\caption{{\it Upper panel}: the number of sources per square degree brighter 
than an X-ray flux limit $S_X$ for the case $\Om=1$. The solid lines correspond 
to galaxies for the ``Hot'' model (upper curve) and the ``Cold'' model (lower 
curve). The dashed line corresponds to groups and clusters. The dotted line 
shows the quasar number counts. We only display the scaling model. The data 
points for quasars are from Hasinger et al. (1998) (disks for the 1112 ksec PSPC 
and triangles for the 207 ksec HRI) and McHardy et al. (1998) (crosses). The data 
points for clusters are as in Fig.\ref{figcountSXclus}. {\it Lower panel}: same 
curves for an open universe with $\Om=0.3$.}
\label{figcountSXgalclus}

\end{figure}

From the X-ray luminosity functions obtained in the previous sections we can 
derive the surface density on the sky of sources brighter than an X-ray flux 
limit $S_X$, taking into account the contribution from galaxies as well as from 
QSOs, groups and clusters. We show our results for the scaling model in 
Fig.\ref{figcountSXgalclus}. First, we note that the prediction of the ``Hot'' 
model for the galaxy counts is much larger than for the ``Cold'' scenario, 
especially for the low-density cosmology. This could be expected from the 
comparison of Fig.\ref{figLXO1} with Fig.\ref{figLXO03}. Thus, in the ``Hot'' 
scenario the contribution from individual galaxies dominates over groups and 
clusters at $S_X \la 10^{-16}$ erg s$^{-1}$ cm$^{-2}$ while in the ``Cold'' 
scenario this only holds for $S_X \la 10^{-18}$ erg s$^{-1}$ cm$^{-2}$ for the 
SCDM case and for even smaller fluxes for the OCDM cosmology. This is due to the 
much lower cutoff $L_{Xf}$ in the OCDM scenario, as seen in  Fig.\ref{figLXO1} 
and Fig.\ref{figLXO03}. Of course, the characteristic fluxes of galaxies are 
much smaller than for clusters because of their smaller mass and temperature. 
Moreover, a significant part of the gas of bright galaxies has been able to cool 
due to their higher redshift of formation. This leads to the sharp high-flux 
cutoff of the galaxy counts as compared with the group and cluster counts, even 
for the ``Hot'' model. However, at low $S_X$ galaxies provide a larger 
contribution than groups because they are much more numerous.

In the ``Cold'' scenario, the observation of these X-ray emitting galaxies would 
provide {\it a direct signature of galaxy formation and of the associated 
cooling process}. Bright galaxies already start to appear at $S_X \sim 10^{-16}$ 
erg s$^{-1}$ cm$^{-2}$, typically one per square degree, and are within the AXAF 
sensitivity limits\footnote{We are indebted to R. Mushotzky for a discussion of 
this point.}. In our model (see Valageas \& Schaeffer 1999), such bright 
galaxies typically correspond to a virial temperature of $10^7 K$ and a baryon 
mass of $ 2.3 \times 10^{11} M_{\odot}$. They are just forming at $z=3$, with an 
X-ray luminosity of $1.5 \times 10^{43}$ erg/s over a time of $1.4 \times 10^9$ 
years. However, in the ``Hot'' scenario most X-ray emitting galaxies would have 
already seen a large part of their gas content cool and the observed 
luminosity would arise from a diffuse hot component, left-over from the last 
merging event or replenished by supernovae, which only contains a small fraction 
of the total baryonic mass. In any case, it is clear that observations (and even 
the lack of galaxy detections at $S_X \sim 10^{-16}$ erg s$^{-1}$ cm$^{-2}$) 
would provide very interesting informations on galaxy formation and evolution. 
As we noted above, however, due to theoretical and observational ambiguities 
the most relevant curve is probably the sum of both contributions from galaxies 
and groups. Thus, we see that the predictions obtained for both cosmologies are 
similar. 

On the other hand, because of their harder radiation spectrum, quasars provide an 
important contribution to the X-ray source counts, which actually dominates over 
the whole range $10^{-18} < S_X < 10^{-12}$ erg s$^{-1}$ cm$^{-2}$. We note that 
although we recover the right abundance of AGN source counts at high 
luminosities $S_X \sim 10^{-14}$ erg s$^{-1}$ cm$^{-2}$ we somewhat 
underestimate the number of low luminosity objects $S_X \sim 10^{-15}$ erg 
s$^{-1}$ cm$^{-2}$. This may suggest that a more detailed model of QSOs is 
needed in order to match exactly the observations. On the other hand, we note 
that McHardy et al. (1998) find that at low fluxes a new population of sources 
appears, which consists of narrow emission lines galaxies which could partly 
correspond to starburst galaxies, which we did not specifically include in our 
model.

\section{Summary and Conclusion}
\label{S&C}

We have examined the predictions of the {\it scaling model} for the number of 
X-ray emitting objects as a function of redshift. This approach gives the 
multiplicity of non-linear structures directly from the non-linear density field and should be considered as an improvement over the PS approximation. This model gives more precise counts and can also be used for any density contrast, both much larger than the usual value of $200$ as well as much smaller. It also solves the hierarchy (cloud-in-cloud) problem and allows us to describe structures embedded within other virialized condensations. Thus we can simultaneously describe 
galaxies (and quasars) as well as galaxy clusters.

For the cluster temperature and X-ray luminosity distributions, the counts 
obtained in this way typically differ by a factor of two from the PS 
approximation, but the differences may reach an order of magnitude for extreme 
cases (very large or very small objects). With an initial CDM spectrum that is 
normalized in the same way as in the numerical simulations that reproduce the 
same data, the scaling model reproduces the currently available observations, 
while the PS approximation does not (the latter would, provided the 
normalization of the power spectrum is modified, with however a different modification depending on the observations to be reproduced).
 The evolution with redshift is 
different too, for reasons that are well understood. The counts for $T > 5$ keV, for instance, peak at $z = 0.4$ rather 
than $z = 0.2$ for $\Om = 1$, with a much larger normalization. 

This allows a more accurate estimation of the Sunyaev-Zel'dovich distorsion 
parameter $y$ along a given line of sight (summing over the contribution of 
clusters). For the same CDM initial conditions as above, we get $\lag y \rag = 2 
\times 10^{-6}$ that induces fluctuations $\lag \delta y^2 \rag^{1/2} = 10^{-6}$ 
in the CMB for $\Om = 0.3$, and values which are a factor of two smaller in a critical 
density universe, the bulk of the contribution arising from clusters at $z < 1$.

Then, we have shown that models of the intra-cluster medium which include a 
characteristic temperature $T_{ad} \sim 0.4$ keV reproduce the main features of 
the observed temperature - X-ray luminosity relation, independently of the 
details of the models. This provides a robust estimate of the luminosity 
function which is seen to show a very weak dependence on redshift.

Next, we have recalled how to  build a model for galaxies and quasars in a 
way that is consistent  with the description of clusters and of the underlying 
density field. In particular, we have pointed out the advantages of our approach 
which allow us to study these high-density objects which cannot be dealt with by 
the standard PS prescription. Thus, we show that the analysis of the observed 
QSO luminosity function in the light of the PS mass function leads to 
discrepancies and to erroneous conclusions due to the intrinsic limitations of 
this theoretical approach. On the other hand, thanks to its more extended range of 
applications, our method can be meaningfully used to study these objects and it 
provides a much better agreement with the data. This also holds for the galaxy 
luminosity function, for the same physical reasons.

Finally, we have 
taken advantage of being able to use 
the same basic model to draw a 
global description of the X-ray emission from all  structures. Thus, we 
find that quasars dominate the X-ray source counts over the available range of 
fluxes. Clusters and groups provide a non-negligible contribution ($\sim 1/5$ of 
the quasar counts) over the same range. On the other hand, at low fluxes $S_X 
\sim 10^{-16}$ erg s$^{-1}$ cm$^{-2}$ one starts to get access to individual 
galaxies, especially for a critical density universe. For the OCDM scenario, 
this only occurs if 
there is inhomogeneous cooling so that
a hot diffuse gaseous component remains in the galactic 
halo while gradually losing mass in the form of cold gas 
(otherwise, if there is no such hot 
gas, galaxies only appear for $S_X \sim 10^{-18}$ erg s$^{-1}$ cm$^{-2}$ in this 
OCDM cosmology). These objects, that were more luminous in the past and start 
appearing in deep enough surveys, are undoubtedly a new challenge. In particular, 
they should be accessible with the current sensitivity of AXAF. Thus, 
observations of X-ray emitting galaxies (or the lack of detection) will provide 
interesting information on galaxy evolution. They will constrain the amount of 
hot gas within galactic halos (which in turn could give some constraints on the 
feedback from supernovae and the infall from the IGM) and especially
 uncover some 
massive galaxies while they are just being formed.

\end{document}